\documentclass{aa}
\usepackage{graphics,psfig}
\begin{document}
   \thesaurus{11    
               (11.06.2; 
                11.16.1; 
                11.19.6; 
                11.19.2
               )
             } 
\title{Three-dimensional modelling of edge-on disk galaxies
\thanks{
Based on observations collected at the European Southern Observatory, Chile 
and Lowell Observatory, Flagstaff (AZ), USA
       }
      }
\author{M. Pohlen \and R.-J. Dettmar \and R. L\"utticke \and U. Schwarzkopf}
\institute{Astronomisches Institut, Ruhr-Universit\"at Bochum, 
           D-44780 Bochum, Germany\\
           email: (pohlen, dettmar, luett, schwarz)@astro.ruhr-uni-bochum.de
          }
\date{Received 13 July 1999 / Accepted 27 March 2000}
\maketitle
   \begin{abstract}
We present detailed three-dimensional modelling of the stellar luminosity 
distribution for the disks of 31 relatively nearby ($\le$ 110 Mpc) edge-on 
spiral galaxies.
In contrast to most of the standard methods available in the literature 
we take into account the full three-dimensional information of the disk. 
We minimize the difference between the observed 2D-image and an image of 
our 3D-disk model integrated along the line of sight. Thereby we specify the
inclination, the fitting function for the z-distribution of the disk, and the
best values for the structural parameters such as scalelength, 
scaleheight, central surface brightness, and a disk cut-off radius. 
From a comparison of two independently developed methods we conclude, that 
the discrepancies e.g. for the scaleheights and scalelengths are of the order 
of $\approx 10\%$. These differences are not due to the individual method 
itself, but rather to the selected fitting region, which masks the bulge 
component, the dust lane, or present foreground stars. Other serious
limitations are small but appreciable intrinsic deviations of 
real disks compared to the simple input model. \\
In this paper we describe the methods and present contour plots as well as
radial profiles for all galaxies without previously published surface 
photometry. Resulting parameters are given for the complete sample. 
\keywords{galaxies: fundamental parameters -- galaxies: surface photometry -- 
galaxies: structure -- galaxies: spiral}
   \end{abstract}
\section{Introduction}
Global parameters of galactic disks can be used to constrain the 
formation process as well as the evolution of disk galaxies. 
Recently it has become possible to deduce parameters such as scalelength or 
central surface density from numerical or semi-analytical self-consistent 
galaxy formation models (Syer et al. \cite{syer}, Mo et al. \cite{mo}, 
Dalcanton et al. \cite{dal}) and compare the results with observed values. 
Observationally the Hubble Deep Field gives the opportunity to study 
morphological features and simple structural parameters for galaxies even at 
high redshifts (Takamiya \cite{taka}, Marleau \& Simard \cite{hdfmorph}, 
Fasano \& Filippi \cite{fasan}). \\
Former statistical studies providing sets of homogeneously derived parameters 
for nearby galaxy samples are those of de Jong (\cite{dj}) using a 
two-dimensional and of Courteau (\cite{cour}) with a one-dimensional 
decomposition technique. \\
However, so far only a few statistical studies based on high quality CCD data 
(de Grijs \cite{dgmn}, Barteldrees \& Dettmar \cite{bd}, hereafter Paper I) 
have addressed the actual three-dimensional structure of disk galaxies with 
regard to the stellar distribution perpendicular to the plane (z-direction) 
and taking into account an outer truncation {\it (cut-off radius)} for the disk, 
first introduced by van der Kruit \& Searle (\cite{vdk81a}). \\
While the stellar distribution perpendicular to the plane (z-profile) could 
result from various "heating" processes during the galactic evolution 
(Toth \& Ostriker \cite{toth}, Hernquist \& Mihos  \cite{hern}), 
the cut-off radius can be used to constrain either the angular momentum of a 
protocloud (van der Kruit \cite{vdk87}) or possible starformation thresholds 
in the gaseous disk (Kennicutt \cite{kenni}). \\
In the following, we have compiled parameters of galactic disks for a total of
31 edge-on galaxies in different optical filters of which 17 objects have been
already discussed in Paper I. One goal of our detailed comparison of several 
independent fitting procedures is to study the influence of the applied 
techniques on resulting disk parameters. It also provides the data base for
statistical analysis addressing some of the above mentioned questions (e.g.,
Pohlen et al. \cite{pol}). For 14 objects 
surface photometric data are given for the first time in Appendix A. 
\section{Observations and data reductions}
\subsection{Observations}
The observations were carried out at the 42-inch (1.06m) telescope of the 
Lowell Observatory located on the Anderson Mesa dark side during several 
nights in December 1988 (run identification: L1) and January 1989 (L2) 
and at the 2.2m telescope at ESO/La Silla during three runs in June 1985 (E1), 
March (E2) and June (E3) 1987. 
At the 42-inch telescope we used a 2:1 focal reducer with the f/8 secondary, 
equipped with a CCD camera which is based on a thinned TI 800x800 WFPC 1 CCD 
with 15 $\mu$m pixelsize resulting in a field of approximately 9$\arcmin$ with 
a scale of 0.7$\arcsec$pixel$^{-1}$. 
Images were taken with a standard Johnson R filter. 
Observations at the 2.2m telescope were carried out with the ESO CCD adapter 
using a 512x320 RCA chip, giving an effective field size of 
$\approx 3\arcmin\;$x$\;2\arcmin$ and a scale of 0.36$\arcsec$pixel$^{-1}$. 
For the ESO observing runs we used the g, r, and i filters of the Thuan and 
Gunn (\cite{tg}) system. Exposures were mainly taken in the g or r band, and 
only seven galaxies were observed in all three filters.
\subsection{Sample selection}
The northern sample observed at Lowell was selected automatically in an 
electronic version of the UGC-catalog (Nilson \cite{ugc}) searching for 
galaxies with an inclination class {\it 7} matching the field size. After 
visual inspection to check the inclination and remove interacting and 
disturbed  galaxies, the observed sample was chosen out of the remaining 
galaxies according to the allocated observing time. \\
For the southern sky there is no comparable catalog providing information 
of inclinations directly. 
Using the axial ratios given e.g. in the ESO-Lauberts \& Valentijn catalog 
(\cite{lv}) will introduce a selection bias preferring late type galaxies with 
lower B/D ratio (Guthrie \cite{guthri}, Bottinelli et al. \cite{bot}). 
One way to avoid this is extending the first selection to much lower axis 
ratios, comparable to $i \approx 65\degr$, and then checking the 
inclination by eye. Therefore we selected the galaxies according to the field 
size of about $2\arcmin$ from a visual inspection of film copies of the southern sky survey (see Paper I). \\
Table \ref{galaxies} gives a list of the resulting sample used during our 
fitting process, with a serial number {\scriptsize{\it (1)}}, 
the principal galaxy name {\scriptsize{\it (2)}}, the used filter 
{\scriptsize{\it (3)}}, the integration time in minutes {\scriptsize({\it 4})},
and the run identification label {\scriptsize{\it (5)}}, whereas the '$\star$' 
marks images already published in Paper I.   
Further parameters are taken from the RC3 catalogue (de Vaucouleurs et al. 
\cite{rc3}): the right ascension {\scriptsize{\it (6)}} and declination 
{\scriptsize{\it (7)}}, the RC3 coded Hubble-type {\scriptsize{\it (8)}}, the 
Hubble parameter T {\scriptsize{\it (9)}}, and the D$_{25}$ diameter in 
arcminutes {\scriptsize{\it (10)}}.  In the case of ESO 578-025 
parameters are taken from the ESO-Uppsala catalogue (Lauberts \cite{eso}).  
   \begin{table*}
      \caption[]{Global parameters of sample galaxies with models}
\begin{tabular}{r l r c l c c l  r@{.}l  r@{.}l  c  r@{.}l }
\hline
No.&name&filter&$t_{int.}$&\multicolumn{1}{c}{run} &RA &DEC&RC3
&\multicolumn{2}{c}{T} 
&\multicolumn{2}{c}{D$_{25}$}
&$v_{\sun}$
&\multicolumn{2}{c}{D} \\
   &    &      &[min]&\multicolumn{1}{c}{ID}  &\multicolumn{2}{c}{(2000.0)}&type &
\multicolumn{2}{c}{}&\multicolumn{2}{c}{[$\;\arcmin\;$]}&km s$^{-1}$
&\multicolumn{2}{c}{Mpc} \\ 
\rule[-3mm]{0mm}{5mm}{\scriptsize{\raisebox{-0.7ex}{\it (1)}}}
&{\scriptsize{\raisebox{-0.7ex}{\it (2)}}}
&{\scriptsize{\raisebox{-0.7ex}{\it (3)}}}
&{\scriptsize{\raisebox{-0.7ex}{\it (4)}}}
&\multicolumn{1}{c}{{\scriptsize{\raisebox{-0.7ex}{\it (5)}}}}
&{\scriptsize{\raisebox{-0.7ex}{\it (6)}}}
&{\scriptsize{\raisebox{-0.7ex}{\it (7)}}}
&{\scriptsize{\raisebox{-0.7ex}{\it (8)}}}
&\multicolumn{2}{c}{{\scriptsize{\raisebox{-0.7ex}{\it (9)}}}}
&\multicolumn{2}{c}{{\scriptsize{\raisebox{-0.7ex}{\it (10)}}}}
&{\scriptsize{\raisebox{-0.7ex}{\it (11)}}}
&\multicolumn{2}{c}{{\scriptsize{\raisebox{-0.7ex}{\it (12)}}}} \\
\hline\hline \\[-0.2cm]
1 &ESO 112-004 &r&40&E3$\star$ &002804.2&$-$580611&  .S.R6*.&5&6   &1&32 & ---& \multicolumn{2}{c}{---} \\
2 &ESO 150-014 &r&20&E3$\star$ &003637.9&$-$565424&  .L..+*/&-0&7 &1&91  &8257&107&05\\
3 &NGC 585   &R&20&L1        &013142.5&$-$005555&  .S..1*/&1&0   &2&14 &5430&72&15\\
4 &ESO 244-048 &r&15&E3        &013908.8&$-$470742&  .S..3./&3&0   &1&38 &6745 &87&09\\
5 &NGC 973   &R&10&L1        &023420.2&+323020&  .S..3..&3&0   &3&72   &4853 &66&28\\
6 &UGC 3326  &R&30&L1        &053936.0&+771800&  .S..6*.&6&0   &3&55   &4085&57&82\\
7 &UGC 3425  &R&30&L1        &061442.0&+663400&  .S..3..&3&0   &2&51   &4057&57&04\\
8 &NGC 2424   &R&15&L1        &074039.8&+391359&  .SBR3*/&3&0   &3&80   &3113&43&10\\
9 &IC 2207    &R&10&L2        &074950.8&+335743&  .S..6*.&6&0   &2&04   &4793&65&10\\
10&ESO 564-027 &r&30&E2        &091154.4&$-$200703&  .S..6*/&6&3   &4&07 &2177&26&93\\
11&ESO 436-034 &g&60&E3$\star$ &103244.2&$-$283646&  .S..3./&3&0   &2&09 &3624&46&02\\
12&ESO 319-026 &g&30&E3$\star$ &113020.0&$-$410357&  .S..5./&5&3   &1&48 &3601&45&32\\
12&ESO 319-026 &i&30&E3  \\  
12&ESO 319-026 &r&30&E2    \\
12&ESO 319-026 &r&30&E3$\star$  \\
13&ESO 321-010 &g&30&E3$\star$ &121142.2&$-$383253&  .S..0*/&0&0   &1&86 &3147&39&53\\
13&ESO 321-010 &r&30&E3$\star$ \\
14&NGC 4835A  &r&40&E3$\star$ &125713.6&$-$462243&  .S..6*/&6&0   &2&57 &3389&42&54\\
15&ESO 575-059 &r&15&E2        &130744.5&$-$192348&  .LA.+?/&-0&8  &1&86 &4570&59&78\\
16&ESO 578-025 &g&30&E2        &140815.5&$-$200019& \multicolumn{1}{c}{---} &1&0   &1&60 & 6364 &83&93\\
16&ESO 578-025 &g&30&E3$\star$  \\           
16&ESO 578-025 &i&30&E2    \\         
16&ESO 578-025 &r&30&E2    \\         
17&ESO 446-018 &r&30&E2       &140838.7&$-$293412&  .S..3./&3&0   &2&34     &4774& 62&15\\
18&IC 4393    &r&30&E2       &141749.5&$-$312056&  .S..6?/&6&0   &2&40     &2754& 35&15\\
19&ESO 581-006 &r&30&E3$\star$    &145803.1&$-$192329&  .SBS7P/&7&0   &1&70 &3119& 40&91\\
20&ESO 583-008 &r&30&E3    &155750.5&$-$222947&  .S?....&6&0   &1&51        &7399& 97&98\\
21&UGC 10535  &r&25&E2    &164600.0&+062800&  .S..2..&2&0  &1&10           &7586&102&48\\
22&NGC 6722   &r&10&E3$\star$    &190339.6&$-$645341&  .S..3./&3&0   &2&88 &5749& 73&79\\
23&ESO 461-006 &r&60&E3$\star$    &195155.9&$-$315852&  .S..5./&5&0   &1&62 &5949& 78&26\\
24&IC 4937    &g&20&E1    &200518.0&$-$561520&  .S..3./&3&0   &1&95        &2337& 28&68\\
24&IC 4937    &i&20&E1       \\	     
24&IC 4937    &r&30&E3$\star$     \\	     
25&ESO 528-017 &g&30&E3$\star$    &203320.8&$-$270549&  .SB.6?/&5&7   &1&59 &6115& 80&73\\
25&ESO 528-017 &i&30&E3   \\	     
25&ESO 528-017 &r&60&E3$\star$  \\	     
26&ESO 187-008 &r&30&E3$\star$    &204325.2&$-$561217&  .S..6./&6&0   &1&51 &4412& 56&31\\
27&ESO 466-001 &i&40&E3    &214232.3&$-$292210&  .S..2./&2&0   &1&38        &7068& 93&13\\
28&ESO 189-012 &g&60&E3$\star$    &215538.7&$-$545233&  .SA.5*/&5&0  &1&66  &8398&109&04\\
28&ESO 189-012 &i&20&E3    \\	     
28&ESO 189-012 &r&30&E3$\star$  \\	     
29&ESO 533-004 &r&20&E1    &221403.2&$-$265618&  .S..5*/&4&8   &2&34        &2594& 33&54\\
30&IC 5199    &g&30&E3$\star$    &221933.0&$-$373201&  .SA.3*/&3&0   &1&55 &5061& 65&78\\
30&IC 5199    &i&30&E3    \\ 	     
31&ESO 604-006 &r&30&E3$\star$    &231454.0&$-$205944&  .S..4./&4&0  &1&86  &7636&100&92\\
\hline
\multicolumn{14}{l}{{\scriptsize $\star$ marks images already published in 
Paper I}}\\
\end{tabular}
         \label{galaxies}
  \end{table*}
\subsection{Data reduction}
We applied standard reduction techniques for bias subtraction, bad pixel 
correction and flatfielding. Following the procedure described in Paper I
the sky background was fitted for each image using the edges of the 
individual frames to reduce any large scale inhomogeneity in the field of view.
For part of the data we tried to remove the foreground stars from the image, 
but even with sophisticated PSF fitting using IRAF-DAOphot routines we were 
not able to remove stars without any confusion. The remaining residuals 
were always of the order of the discussed signal. This technique could only 
be used to mask out the regions affected by stars. In order to increase the 
signal-to-noise ratio near the level of the sky background part of the data 
was filtered using a weighted smoothing algorithm (see Paper I). 
Thereby the noise was reduced by about one magnitude measured 
with a three sigma deviation on the background, whereas the interpretation of 
the faint structure is hampered by this process.  \\
We therefore conclude that the best way is to omit any additional image 
modifications, besides a rotation of the image to the major axis of the disk.
\subsection{Photometric calibration}
\label{calib}
Most of the images were taken during non-photometric nights, therefore we 
tried a different way to perform photometric calibration. Comparing simulated 
aperture measurements with published integrated aperture data led to the
best possible homogeneous calibration of the whole sample. 
Most of the southern galaxies were calibrated using the catalogue of 
Lauberts \& Valentijn (\cite{lv}), whereas NED\footnote{NASA/IPAC 
Extragalactic Database (NED)} was used for all northern galaxies. 
We used equation (1) derived in Paper I for the colour transformation of R 
and B literature values and the g measurements, and no correction between the 
R and r, and  I and i band, respectively. 
Due to the fact that the photometric errors, of the input catalogues from
Lauberts \& Valentijn (\cite{lv}) as well as within the RC3 catalogue 
(de Vaucouleurs et al. \cite{rc3}) are of the order of $0.1$ mag, we do not 
apply any further corrections.   
Galaxies calibrated in this way are marked with {\it l} in column 
{\scriptsize{\it (5)}} of Table \ref{modelle}, whereas for galaxies which did 
not have published values for their magnitude in the observed filter, we 
interpolated from calibrated images of the same night, by comparing the 
count rates for the sky value. These galaxies are marked with an {\it i}. 
For a few nights no galaxy with published photometry was observed and in these
cases we used interpolated night sky values from the same observing 
run. Images calibrated in this way are marked with {\it e} in Table 
\ref{modelle}.  
The resulting zero points and central surface brightness 
values in these cases should be interpreted carefully, although the derived 
structural parameters, like scalelength and scaleheight, are not  
influenced by any uncertainty in the flux calibration. \\
Appendix A shows the contour plots and selected radial profiles 
for the 14 objects (25 images) not already published in Paper I.
\subsection{Distance estimates}
In order to derive the intrinsic values of the scale parameters and to compare 
physical dimensions we tried to estimate distances for our galaxies. 
Therefore we took published radial velocities corrected for the Virgo 
centric infall from the LEDA\footnote{Lyon/Meudon Extragalactic 
Database (LEDA)} database, and estimated the distance following the Hubble 
relation 
with a Hubble constant of $H_{0}\!=\!75$ km s$^{-1}$Mpc$^{-1}$. 
Table \ref{galaxies} gives the heliocentric radial velocities 
{\scriptsize{\it (11)}} according to LEDA, and our estimated distances 
{\scriptsize{\it (12)}}.
%
\section{Disk models}
\subsection{Background}
Our disk model is based upon the fundamental work of van der Kruit and Searle 
(\cite{vdk81a}, \cite{vdk81b}, \cite{vdk82a}, \cite{vdk82b}). 
They tried to find a fitting function for the light distribution in
disks of edge-on galaxies. 
These galaxies are, compared to the face-on view, preferred for studying 
galactic disks due to the fact, that in this geometry it is possible to 
disentangle the radial and vertical stellar distribution.
Their model include an exponential radial light distribution found for 
face-on galaxies (de Vaucouleurs, \cite{devau59}; Freeman, \cite{free}), a 
sech$^2$ behaviour in $z$, which is expected for an isothermal population in a 
plan-parallel system (Camm, \cite{camm};  Spitzer, \cite{spitzer}), and a 
sharp edge of the disk, first observed by van der Kruit (\cite{vdk79}) in 
radial profiles of edge-on galaxies. The resulting luminosity density 
distribution for this symmetric disk model is 
(van der Kruit \& Searle \cite{vdk81a}):
\begin{equation}
\hat{L}(R,z) = \hat{L}_0 \ \exp{\left(-\frac{R}{h}\right)} \ 
{\rm sech}^2{\left(\frac{z}{z_0}\right)} \qquad R < R_{\rm co}
\end{equation}
$\hat{L}$ being the luminosity density in units of [$L_{\sun}$ pc$^{-3}$],
$\hat{L}_0$ the central luminosity density, $R$ and $z$ are the radial resp. 
vertical axes in cylinder coordinates, $h$ is the radial scalelength and 
$z_0$ the scaleheight, and $R_{\rm co}$ is the cut-off radius. \\
The empirically found exponential radial light distribution is now well 
accepted and it is proposed that viscous dissipation could be responsible
(Firmani et al. \cite{firmani}, Struck-Marcel \cite{struck}; 
Saio \& Yoshii \cite{saio}, Lin \& Pringle \cite{lp}), although there is so 
far no unique explanation for the disk being exponential.
An alternative description of the form $1/R$ proposed by Seiden et al. 
(\cite{seiden}) did not get much attention, although it emphasizes the 
empirical nature of the exponential fitting function. \\
To avoid the strong dust lane and to follow the light distribution down to 
the region $z\!\approx\!0$ Wainscoat (\cite{wain}) and Wainscoat et al. 
(\cite{wainetal}) carried out NIR observations using the much lower 
extinction in this wavelength regime compared to the optical. 
They found a clear excess over the 
isothermal distribution and proposed the z-distribution to be better fitted by 
an exponential function $f(z)\!=\!\exp (-z/z_{0})$.   
According to van der Kruit (\cite{vdk88}) such a distribution would led to 
a sharp minimum of the velocity dispersion in the plane, which is not observed 
(Fuchs \& Wielen \cite{fw}). Therefore he proposed 
$f(z)\!=\!{\rm sech} (z/z_{0})$
as an intermediate solution. De Grijs (\cite{dg}) extended this to a family 
of density laws 
$g_m(z,z_0)\!=\! 2^{-2/m}\ g_0\ {\rm sech}^{2/m} \left( mz/2 z_0  \right) (m>0)$
following van der Kruit (\cite{vdk88}), where the isothermal ($m=1$), and 
the exponential ($m=\infty$) cases represent the two extremes. \\
Therefore the luminosity density distribution can be written as:
\begin{equation}
\hat{L}(R,z) = \hat{L}_0 \ \exp{\left(-\frac{R}{h}\right)} \ f_n(z,z_0) \ {\rm H}(R_{\rm co}-R)
\label{hatl}
\end{equation}
with H$(x_0-x)$ being the Heaviside function. \\
In order to limit the choice of parameters we restrict our models to the 
three main density laws for the z-distribution (exponential, sech, and 
sech$^2$). Due to the choice of our normalised isothermal case $z_0$ is equal 
to $2 h_z$, where $h_z$ is the usual exponential vertical scale height: 
\begin{eqnarray}
\nonumber
f_1(z) &=& 4\ \exp{\left(-2\ \frac{\mid z \mid}{z_0}\right)} \\
\nonumber
f_2(z) &=& 2\ {\rm sech} \left(\frac{2 z}{z_0} \right) \\
\nonumber
f_3(z) &=& {\rm sech}^2{\left(\frac{z}{z_0}\right)}
\end{eqnarray}
In contrast to Paper I and Barteldrees \& Dettmar (\cite{bdold}) we define 
the cut-off radius at the position where the radial profiles become  nearly 
vertical, corresponding to the mathematical description. They tried to avoid 
any confusion due to the lower signal-to-noise in the outer parts, by fixing 
the cut-off radius where the measured radial profile begin to deviate 
significantly from the pure exponential fit. 
\subsection{Numerical realisation}
\begin{figure*}[t]
\psfig{figure=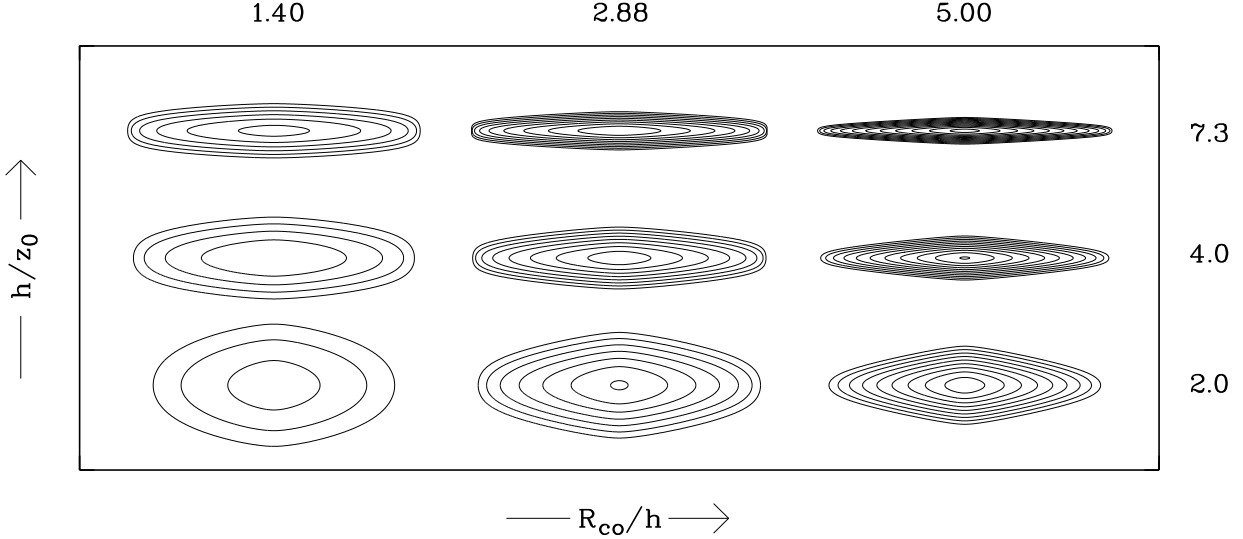,width=18cm,angle=270}
\caption{Different galaxy models computed according to equation 
(\ref{numcalhatl})
with constant values for the ratio $\frac{R_{\rm co}}{h}$ within the columns, or 
$\frac{h}{z_0}$ within the rows, whereas $\hat{L}_{tot}$ and $R_{\rm co}$ are
constant for all models using an isothermal description for the exact edge-on 
case}
\label{galacticmodels}
\end{figure*}

The model of the two dimensional surface photometric intensity results from 
an integration along the line of sight of the three dimensional luminosity 
density distribution (\ref{hatl}) with regard to the inclination $i$ of the 
galaxy. Describing the luminosity density of the disk in a 
cartesian-coordinate grid K($x$-$y$-$z$) with $R=\sqrt{(x^2+y^2)}$ leads to 
the following coordinate transformation into the observed inclined system 
K$'$($x'$-$y'$-$z'$) with $x'$ pointing towards the observer, whereas the rotation angle 
between the two systems corresponds to  $90\degr-i$:
\begin{eqnarray*}
x &=& x' \sin(i) - z'\cos(i)\\
y &=& y'\\
z &=& x' \cos(i) + z'\sin(i) 
\end{eqnarray*}
Taking into account this transformation we have to integrate equation 
(\ref{hatl}), obtaining an equation for the intensity of the model disk 
depending on the observed radial and vertical axes $y'$ and $z'$ on the CCD:
\begin{equation}
I(y',z') 
= \int\limits_{-\infty}^{+\infty} 
\hat{L}\bigl(x(x',z',i), y', z(x',z',i)\bigr) dx' 
\label{loshatl}
\end{equation}
Together with equation (\ref{hatl}) this gives:
\begin{eqnarray}
\nonumber 
I(y',z') = \qquad  \qquad\qquad\qquad\qquad\qquad\qquad\qquad\qquad \\[0.3cm]
\nonumber
\hat{L}_0 \int\limits_{-\infty}^{+\infty}\!\!
\exp{\left(-\frac{\sqrt{(x'\ \sin i - z'\ \cos i)^2+y'^2}}{h}\right)} f_n(z',z_0)\\
\ast\, {\rm H}\left(R_{\rm co}-\sqrt{(x'\ \sin i - z'\ \cos i)^2+y'^2}\right) \:dx' \quad
\label{numcalhatl}
\end{eqnarray}
Therefore six free parameters fit the observed surface intensity on the chip 
($y'$, $z'$ plane) to the model:
\begin{equation}
I = I(y',\ z',\ i,\ n,\ \hat{L}_0,\ R_{\rm co},\ h,\ z_0)
\label{param}
\end{equation}
Figure \ref{galacticmodels} shows a sequence of computed models with an 
isothermal z-distribution ($n\!=\!3$) for the exact edge-on case
($i\!=\!90\degr$) with characteristic values for the ratio $R_{\rm co}/h$: 
$1.40,2.88,5.00$ and for $h/{z_0}$: $2.0,4.0,7.3$ keeping the cut-off radius
$R_{\rm co}$ and the total luminosity $L_{tot}\!\propto\!z h^2 \hat{L}_0$
constant. The latter is causing a different central surface brightness 
$\mu_{0}$ for each model, ranging from 23.8 to 19.3 starting with  
$\mu_0\!=\!21.2$ mag$/\sq\arcsec$ for the reference model with $R_{\rm co}/h\!=\!2.88$
and $h/{z_0}\!=\!4.0$. All contour lines falling within the interval 
$\mu\!=\!25.0$ -- $19.5$ are plotted with a spacing of 0.5.  
\subsection{Method 1}
\subsubsection{Determination of fitting area}
The first step is to divide the galaxy into its quadrants. The four images are
then averaged according to their orientation, following van der 
Kruit \& Searle (\cite{vdk81a}) and  Shaw (\cite{s93}). 
Thereby larger foreground stars and asymmetrical perturbations in the 
intensity distribution are eliminated by omitting this region during averaging.
Smaller foreground stars are removed by median filtering. The average 
quadrant should result at least from three quadrants to get a representative 
image of the galaxy. This averaging additionally increases the signal-to-noise
ratio. 
In order to determine the fitting area for modelling the disk component on the
final quadrant, one has to avoid the disturbing influence of the bulge 
component and the dust lane. 
The region dominated by the bulge is fixed following Wyse et al. (\cite{wgf}) 
defining the bulge component by ``light that is in excess of an inward 
extrapolation of a constant scale-length exponential disk''. Therefore the 
clear increase of the intensity towards the center which can be seen in 
radial cuts determines an inner fitting boundary $R_{\rm min}$. 
We tried to minimize the dust influence (cf. Section \ref{dust}) by placing 
a lower limit $z_{\rm min}$ in the vertical direction by visual inspection.
Additionally we restricted the remaining image by a limiting contour line
$\mu_{\rm lim}$, where the intensity drops below a limit of 3$\sigma$ on the 
background. \\
We are aware of the problem that these are rather rough definitions difficult
to reproduce without quoting the exact values for $R_{\rm min}, z_{\rm min}, 
\mu_{\rm lim}$.
However, the final choice of the fitting area is a complex and subjective 
procedure depending on the intrinsic shape of each individual galaxy, the 
influence of their environment, and the quality of the image itself.     
Therefore it is not possible to quote exact general selection criteria and
to derive the structural parameters straight forward.
One solution is to do it in a consistent way for a large sample, 
leaving the problem of comparing results from different methods 
(cf. Section \ref{comp}). 
\subsubsection{Numerical fitting}
The numerical realisation of the fitting procedure minimizes the difference 
($SQ$) between the averaged quadrant and a 
modelled quadrant based on equation (\ref{numcalhatl}).
\begin{displaymath}
\nonumber
SQ = \sum_{j}\Bigl(log\left(I_{O_j}(y_j,z_j)\right)-log\left(I_{M_j}(y_j,z_j)\right)\Bigr)^2 
\end{displaymath}
$I_{O_j}$ is the intensity within the average quadrant (observed intensity) 
and $I_{M_j}$ of the modelled intensity. 
In contrast to Shaw \& Gilmore (\cite{sg89}, \cite{sg90}) and 
Shaw (\cite{s93}) using a similar approach for their models we do not weight 
individual pixels. They weight the difference by the error in the surface 
brightness measure, which Shaw (\cite{s93}) derives from the averaging of 
the quadrants. However, this method implies an absolute symmetry for the disk
in $z$ and $y$, which is not the case for real galaxies. 
These kind of errors only reflect the asymmetry of the galaxy. Using the 
observed errors in the surface brightness for weighting individual pixels 
does not result in a considerable 
advantage because they are nearly the same after smoothing.
The minimal $SQ$ is found by varying five of the six free parameters of the 
model (cf. eq. \ref{param}), whereas the parameter $R_{\rm co}$ is determinated 
by cuts parallel to the major axis. A significant decrease of the intensity 
extrapolated to $I = 0$ gives the value of $R_{\rm co}$ (van der Kruit \& Searle, 
\cite{vdk81a}), therefore it is important that the intensity at the cut-off 
radius is well above the noise limit. 
The other five parameters are determined by fixing the smallest $SQ$. 
For the three different functions $f(z)$ and every possible $i$ 
($\Delta i = 0.5\degr$) the remaining parameters ($\hat{L}_0$, $h$, and $z_0$)
are varied with the ``down\-hill simplex-method'' (Nelder \& Mead, \cite{nm}; 
Press et al., \cite{pftv}) until  the global minimum of $SQ$ is found.
$I_{M_j}$ is calculated by a numerical gaussian integration of equation 
(\ref{numcalhatl}). The possible inclination angles can be restricted from 
the dust lane of the galaxy (Paper I). 
Tests of model disks with additional noise show that the ``down\-hill 
simplex-method'' found the input inclination $i$, the used model
$f(z)$ and the other disk parameters within errors of $\delta h, \delta z_0$, 
and $\delta \hat{L}_0 <$ 1\%. 
An estimation of the errors of the parameters for the best model disk can be 
made by inspection of the parameter space for $SQ$ around the smallest $SQ$, 
with slightly different fitting areas, and different values of $R_{\rm co}$. 
$f(z)$ is in almost all cases the same and the variation of $i$ is 
only small ($\pm 1\degr$). The differences in $h$ and $z_0$ are about 
15\% (in some cases up to 25\%), and $\hat{L}_0$ varies about a factor of 2.
%
%
\subsection{Method 2}
\begin{figure}[t]
\psfig{figure=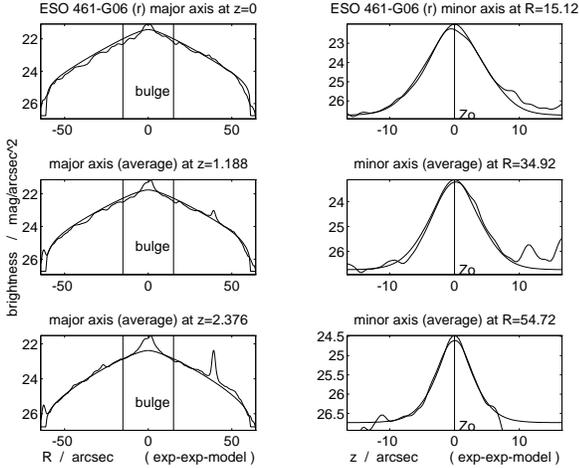,width=9cm,angle=0}
\caption{Example of the fitting procedure: Set of typical radial
(left panel) and vertical (right panel) profiles used for the fitting process
of the galaxy ESO 461-006 together with the model.} 
\label{uwemod}
\end{figure}
Within this method all disk parameters as well as the choice of the 
optimum function $f(z)$ are estimated by a direct comparison of calculated 
and observed disk profiles by eye. \\
As a first step, the inclination $i$ is determined by using the axis
ratio of the dust lane. Depending on the shape of the dust lane it is 
possible to restrict the inclination to $\pm 1.5 \degr$. The central 
luminosity density $\hat{L}_0$ is calculated automatically for each new 
parameter set by using a number of preselected reference points along the 
disk. Given a sufficient signal-to-noise ratio, the cut-off radius $R_{\rm co}$ can
be determined from the major axis profile. Thus, the remaining fitting 
parameters are the disk scale length and height, $h$ and $z_0$, as well as 
the set of 3 functions $f(z)$. 
The scale length $h$ is fitted to a number (usually between 4 and 6) of major 
axes profiles (left panel of Fig. \ref{uwemod}).
The quality of the fit for the vertical profiles along the disk can be used 
as a cross-check for the estimated scalelengths (right panel of 
Fig. \ref{uwemod}).
The disk scaleheight $z_0$ is estimated by fitting the $z$-profiles, outside 
a possible bulge or bar contamination. 
The vertical disk profiles of most of the galaxies investigated enable
a reliable choice of the quantitatively best fitting function $f(z)$.
This is due to the fact that the deviations between different functions 
become visible at vertical distances larger than that of the most 
sharply-peaked dust regions. 
The first raw fitting steps are usually carried out by using a reduced
number of both major and minor axis profiles simultaneously. Afterwards, 
when a good first fitting quality is reached, a complete set of major and 
minor calculated and observed axes profiles are investigated in detail. 
At the end of this procedure a final, complete disk model is calculated 
using all previously estimated disk parameters. 
\section{Results}
\subsection{Distribution of disk parameters}
Table \ref{modelle} contains the best fit model for each image. Together with 
the galaxy name {\scriptsize{\it (1)}}, the filter {\scriptsize{\it (2)}}, 
and the referring image {\scriptsize{\it (3)}}, with integration time, 
and run ID, we list the inclination {\scriptsize{\it (4)}}, the best fitting 
function for the z-distribution {\scriptsize{\it (5)}}, the calibration index 
{\scriptsize{\it (6)}} (ref. Section \ref{calib}), and the central surface 
brightness of the model {\scriptsize{\it (7)}}, without correcting for 
inclination. According to the distance tabulated in Table \ref{galaxies},
the cut-off radius $R_{\rm co}$ {\scriptsize{\it (8)}} is given in kpc and arcsec
as well as the scalelength $h$ {\scriptsize{\it (9)}} and the vertical 
scaleheight $z_0$ {\scriptsize{\it (10)}} which is normalised to the isothermal 
case, being two times an exponential scaleheight $h_z$. 
For the seven galaxies with available images in more than one filter, we 
do not see any correlation of fitted parameters with different wavelength,
although we find the same inclination angle for the best fitted disk within 
the range of the errors. 
Appendix A shows the best fitting model as an overlay to selected 
radial profiles for each image. The subsequent analysis of the distribution 
for the different parameters concerning the formation and evolution of 
galaxies will be given in forthcoming papers.
   \begin{table*}
      \caption[]{Determined parameters set}
\begin{tabular}{l c c r@{.}l lcc r@{.}l r@{.}l cr@{.}lr@{.}l cr@{.}lr@{.}l}
\hline
\hspace*{0.5cm} galaxy&filter&image&\multicolumn{2}{c}{$i$}&$f(z)$&cali.&$\mu_0$
&\multicolumn{4}{c}{$R_{\rm co}$}&
&\multicolumn{4}{c}{$h$} &
&\multicolumn{4}{c}{$z_0$} \\
\cline{9-12}\cline{14-17}\cline{19-22} 
&&&\multicolumn{2}{c}{\raisebox{-0.7ex}{[$\degr$]}}&&& mag$/\sq\arcsec$
&\multicolumn{2}{l}{\raisebox{-0.7ex}{[$\;\arcsec\;$]}}&\multicolumn{2}{r}{\raisebox{-0.7ex}{[kpc]}}&
&\multicolumn{2}{l}{\raisebox{-0.7ex}{[$\;\arcsec\;$]}}&\multicolumn{2}{r}{\raisebox{-0.7ex}{[kpc]}}&
&\multicolumn{2}{l}{\raisebox{-0.7ex}{[$\;\arcsec\;$]}}&\multicolumn{2}{r}{\raisebox{-0.7ex}{[kpc]}} \\
\hspace*{0.8cm}\rule[-3mm]{0mm}{5mm}{\scriptsize{\raisebox{-0.7ex}{\it (1)}}}
&{\scriptsize{\raisebox{-0.7ex}{\it (2)}}}
&{\scriptsize{\raisebox{-0.7ex}{\it (3)}}}
&\multicolumn{2}{c}{{\scriptsize{\raisebox{-0.7ex}{\it (4)}}}}
&\hspace*{0.15cm}{\scriptsize{\raisebox{-0.7ex}{\it (5)}}}
&{\scriptsize{\raisebox{-0.7ex}{\it (6)}}}
&{\scriptsize{\raisebox{-0.7ex}{\it (7)}}}
&\multicolumn{4}{c}{{\scriptsize{\raisebox{-0.7ex}{\it (8)}}}}&
&\multicolumn{4}{c}{{\scriptsize{\raisebox{-0.7ex}{\it (9)}}}} &
&\multicolumn{4}{c}{{\scriptsize{\raisebox{-0.7ex}{\it (10)}}}} \\
\hline\hline \\[-0.2cm]
ESO 112-004 &  r&40E3  &  87&5 & sech    &l& 21.14  &  45&0  & \multicolumn{2}{c}{---}  &&  22&7  &    \multicolumn{2}{c}{---}  & & 2&8  &   \multicolumn{2}{c}{---}   \\
ESO 150-014 &  r&20E3  &  90&0 & sech    &l& 22.00  &  64&1  &33&27 &&  23&4  &12&14 & & 5&5  &2&85\\
NGC 585   &  R&20L1   &  88&0 & sech    &e& 21.76  &  70&0  &24&49 &&  32&7  &11&44 & &10&6  &3&71\\
ESO 244-048 &  r&15E3  &  87&0 & $\exp$  &l& 20.54  &  45&4  &19&17 &&  13&7  & 5&78 & & 7&2  &3&04\\
NGC 973   &  R&10L1   &  89&5 & $\exp$  &e& 20.83  & 105&0  &33&74 &&  51&5  &16&55 & &12&4  &3&99\\
UGC 3326  &  R&30L1   &  88&0 & sech$^2$&e& 21.50  & 101&5  &28&45 &&  70&4  &19&74 & & 4&7  &1&32\\
UGC 3425  &  R&30L1   &  87&0 & sech$^2$&e& 21.01  &  80&5  &22&26 &&  29&2  & 8&08 & & 7&4  &2&05\\
NGC 2424  &  R&15L1    &  86&5 & $\exp$  &e& 20.52  & 112&0  &23&40 &&  31&7  & 6&62 & &11&9  &2&49\\
IC  2207   &  R&10L2   &  86&5 & $\exp$  &e& 21.04  &  56&0  &17&67 &&  38&2  &12&06 & & 6&7  &2&12\\
ESO 564-027 &  r&30E2  &  88&0 & sech    &l& 21.11  & 140&4  &18&33 &&  50&9  & 6&65 & & 6&6  &0&86\\
ESO 436-034 &  g&60E3  &  88&0 & sech    &l& 20.96  &  82&1  &18&32 &&  22&6  & 5&04 & & 6&8  &1&52\\
ESO 319-026 &  g&30E3  &  86&5 & sech    &e& 21.51  &  60&1  &13&21 &&  14&2  & 3&12 & & 3&1  &0&68\\
ESO 319-026 &  r&30E2  &  86&0 & $\exp$  &i& 21.14  &  63&0  &13&84 &&  14&7  & 3&23 & & 3&2  &0&70\\
ESO 319-026 &  r&30E3  &  88&0 & sech$^2$&i& 21.94  &  61&9  &13&60 &&  13&2  & 2&90 & & 2&8  &0&62\\
ESO 319-026 &  i&30E3  &  88&0 & sech    &l& 21.01  &  64&8  &14&24 &&  14&3  & 3&14 & & 3&4  &0&75\\
ESO 321-010 &  g&30E3  &  88&0 & sech    &l& 20.11  &  64&4  &12&34 &&  19&9  & 3&81 & & 4&7  &0&90\\
ESO 321-010 &  r&30E3  &  88&0 & sech    &l& 19.54  &  64&8  &12&42 &&  21&6  & 4&14 & & 5&0  &0&96\\
NGC 4835A  &  r&40E3   &  85&5 & $\exp$  &l& 20.92  &  90&0  &18&56 &&  40&4  & 8&33 & & 7&9  &1&63\\
ESO 575-059 &  r&15E2  &  87&0 & sech$^2$&l& 20.78  &  60&5  &17&53 &&  22&6  & 6&55 & & 6&4  &1&86\\
ESO 578-025 &  g&30E2  &  86&5 & $\exp$  &l& 21.58  &  50&7  &20&63 &&  17&0  & 6&92 & & 7&5  &3&05\\
ESO 578-025 &  g&30E3  &  86&5 & $\exp$  &l& 21.69  &  50&7  &20&63 &&  17&0  & 6&92 & & 7&5  &3&05\\
ESO 578-025 &  r&30E2  &  86&0 & $\exp$  &l& 21.02  &  50&4  &20&51 &&  14&9  & 6&06 & & 7&4  &3&01\\
ESO 578-025 &  i&30E2  &  86&0 & $\exp$  &l& 20.27  &  47&5  &19&33 &&  20&3  & 8&26 & & 7&6  &3&09\\
ESO 446-018 &  r&30E2  &  86&5 & sech    &l& 20.45  &  75&6  &22&78 &&  23&7  & 7&14 & & 3&9  &1&18\\
IC  4393   &  r&30E2   &  87&0 & $\exp$  &l& 20.30  &  75&6  &12&88 &&  29&8  & 5&08 & & 5&8  &0&99\\
ESO 581-006 &  r&30E3  &  86&5 & $\exp$  &l& 21.33  &  55&8  &11&07 &&  18&2  & 3&61 & & 5&6  &1&11\\
ESO 583-008 &  r&30E3  &  87&0 & $\exp$  &i& 20.84  &  56&2  &26&70 &&  14&0  & 6&65 & & 3&8  &1&81\\
UGC 10535  &  r&25E2   &  88&0 & sech$^2$&i& 21.50  &  41&4  &20&57 &&  11&4  & 5&66 & & 4&3  &2&14\\
NGC 6722   &  r&10E3   &  86&5 & sech    &l& 19.47  &  86&0  &30&77 &&  21&6  & 7&73 & & 6&2  &2&22\\
ESO 461-006 &  r&60E3  &  87&5 & $\exp$  &l& 20.96  &  61&6  &23&37 &&  21&1  & 8&01 & & 3&8  &1&44\\
IC  4937   &  g&20E1   &  88&5 & sech$^2$&l& 22.73  &  74&9  &10&41 &&  93&8  &13&04 & & 5&6  &0&78\\
IC  4937   &  r&30E3   &  88&0 & sech$^2$&l& 21.29  &  83&5  &11&61 &&  32&1  & 4&46 & & 5&9  &0&82\\
IC  4937   &  i&20E1   &  88&5 & sech    &e& 19.92  &  78&1  &10&86 &&  38&6  & 5&37 & & 7&2  &1&00\\
ESO 528-017 &  g&30E3  &  86&5 & $\exp$  &l& 21.98  &  57&6  &22&54 &&  20&5  & 8&02 & & 3&5  &1&37\\
ESO 528-017 &  r&60E3  &  86&5 & $\exp$  &l& 21.28  &  55&1  &21&56 &&  19&9  & 7&79 & & 2&7  &1&06\\
ESO 528-017 &  i&30E3  &  86&5 & sech    &l& 20.89  &  50&0  &19&57 &&  22&6  & 8&85 & & 3&2  &1&25\\
ESO 187-008 &  r&30E3  &  85&5 & $\exp$  &l& 20.77  &  50&0  &13&65 &&  15&1  & 4&12 & & 4&4  &1&20\\
ESO 466-001 &  i&40E3  &  87&0 & $\exp$  &e& 19.52  &  52&6  &23&76 &&  13&0  & 5&87 & & 8&2  &3&70\\
ESO 189-012 &  g&60E3  &  87&0 & $\exp$  &l& 21.50  &  56&2  &29&81 &&  26&8  &14&21 & & 3&8  &2&02\\
ESO 189-012 &  r&30E3  &  86&5 & $\exp$  &l& 20.71  &  56&5  &29&97 &&  20&6  &10&93 & & 3&4  &1&80\\
ESO 189-012 &  i&20E3  &  87&0 & sech    &l& 20.44  &  54&7  &29&01 &&  22&7  &12&04 & & 3&2  &1&70\\
ESO 533-004 &  r&20E1  &  88&0 & $\exp$  &l& 20.32  &  68&4  &11&12 &&  33&1  & 5&38 & & 7&7  &1&25\\
IC  5199   &  g&30E3   &  86&5 & $\exp$  &l& 21.55  &  64&1  &20&44 &&  19&9  & 6&35 & & 5&6  &1&79\\
IC  5199   &  i&30E3   &  86&5 & $\exp$  &l& 19.97  &  57&6  &18&37 &&  19&9  & 6&35 & & 4&9  &1&56\\
ESO 604-006 &  r&30E3  &  90&0 & sech    &l& 21.22  &  70&6  &34&54 &&  27&9  &13&65 & & 3&8  &1&86\\
\hline
\end{tabular}
         \label{modelle}
  \end{table*}
\subsection{Comparison of  different methods}
\label{comp}
The different fitting methods were independently developed within two diploma 
theses (L\"utticke 1996, Schwarz\-kopf 1996). 
The quality of the data basis for each project was the same. From the sample 
presented here there were five objects in common. These are used to compare 
the two methods and determine the quantitative difference of the derived 
parameters.
\begin{table}[h]
 \caption[]{Comparison of the different determined parameter sets for the 
            same galaxy images}
 \label{verglRLUS}
\begin{tabular}{r l c c c c cl}
\hline
galaxy&$f(z)$  & $i$ & $z$ & $h$ & $R_{\rm co}$   \\
&
&\raisebox{+0.2ex}[1.0ex]{[$\;\degr\;$]}
&\raisebox{+0.2ex}[1.0ex]{ [$\;\arcsec\;$] }
&\raisebox{+0.2ex}[1.0ex]{ [$\;\arcsec\;$] }
&\raisebox{+0.2ex}[1.0ex]{ [$\;\arcsec\;$] } \\
\hline\hline \\[-0.2cm]
&{\rm sech}&88.0&5.0&21.6&64.8 \\
\raisebox{1.1ex}[1.1ex]{ESO 321-010 r {\large \{}}&$\exp$&88.0&5.2&25.9&64.1 \\
&$\exp$&87.5&3.8&21.1&61.6 \\
\raisebox{1.1ex}[1.1ex]{ESO 461-006 r {\large \{}}&$\exp$&87.5&3.6&16.2&61.2 \\
&$\exp$&86.5&6.1&26.8&83.5 \\
\raisebox{1.1ex}[1.1ex]{IC 4937 r {\large \{}}&{\rm sech}&89.5&7.0&27.4&75.6 \\
&$\exp$&86.5&3.2&19.3&56.2 \\
\raisebox{1.1ex}[1.1ex]{ESO 189-012 r {\large \{}}&$\exp$&88.0&3.6&21.3&56.9 \\
&{\rm sech}      &88.5&4.1&39.8&67.7 \\
\raisebox{1.1ex}[1.1ex]{ESO 604-006 r {\large \{}}&{\rm sech}$^2$&89.5&3.2&24.5&73.4 \\
\hline
\end{tabular}
           \end{table}
Table \ref{verglRLUS} shows the results for the five images. The mean 
deviation in the determined inclination is $\approx \! 1 \degr$ and 12.4\% 
for the scaleheight (ranging from 5.0\%-26.6\%) whereas for three images 
different functions for the z distribution were used. The mean difference for 
the radial scalelength is 20.6\% (2.1\%-47.2\%) and 4.2\% for the 
determination of the cut-off radius. \\
A subsequent analysis shows that it is not possible to ascribe the sometimes 
quite large discrepancies to the quality of the individual method. It turns
out, that the main problem is the non-uniform determination of the fitting 
area. The intrinsic asymmetric variations of a real galactic disk compared
to the model enforce a more subjective restriction of the galaxy image to the 
fitting region, whereby for example one has to exclude the bulge 
area and the dust lane. \\
This finding is in agreement with the study of Knapen \& van der Kruit 
(\cite{knap}) who compared published values of the scalelength and 
find an average value of 23\% for the discrepancy between different sources.
As already mentioned by Schombert \& Bothun (\cite{sb}) the limiting factor
for accuracy of the decomposition is not the typical S/N from the 
CCD-telescope combination nor the errors in the determination of the sky
background, but the deviation of real galaxies from the standard model.    
\subsection{Comparison with the literature}
In our former study (Paper I) with an earlier method to adapt equation 
(\ref{numcalhatl}), 20 of our 45 galaxy images have already been used.
We decided to re-use them in this study to get models for as many galaxies as 
possible in a homogeneous way. Additionally, Paper I only presents the best 
fit values for the isothermal model, and uses a different definition of the 
cut-off radius. \\
Only three galaxies are in common with the sample of de Grijs (\cite{dgmn}): 
ESO 564-027, ESO 321-010, and ESO 446-018. The mean difference for the 
scalelength is 10.4 \% (ranging from 1.2\%-20.3\%) and for the scale height
(normalised to the isothermal case) 4.0\% (0.0\%-8.2\%).
For the remaining galaxies there are no models in the literature.   
\subsection{Model limitations}
Our model only represents a rather simple axisymmetric three-dimensional 
model for a galactic disk, consisting of an one component radial exponential 
disk with three different laws for the density distribution in 
the z-direction and a sharp outer truncation. Therefore it does not include 
additional components, such as bulges, bars, thick disks, or rings, and
cannot deal with any asymmetries. 
Features like spiral structure or warps are not included, whereas 
Reshetnikov \& Combes (\cite{rescom}) multiply their exponential disk by a 
spiral function introducing an expression to characterise an intrinsic warp 
depending on the position angle outside of a critical radius. \\
The choice of our fitting area tries to avoid the dust lane,
possible only for almost edge-on galaxies, as a first step to account for 
the dust influence (cf. Section \ref{dust}). Examples of models including a 
radiative transfer with an extinction coefficient $\kappa_{\lambda}(R,z)$ can 
be found in Xilouris et al. (\cite{xil99}). 
However, introducing more and more new components and features automatically 
increases the amount of free parameters. Therefore we restricted our model to 
the described six parameters, to obtain statistically meaningful 
characteristics for galactic disks. \\
In the following we demonstrate that a simple disk model omitting the bulge 
component and the dust lane give indeed reasonable parameters. 
\subsubsection{The influence of the bulge component}
\begin{figure}[t]
\psfig{figure=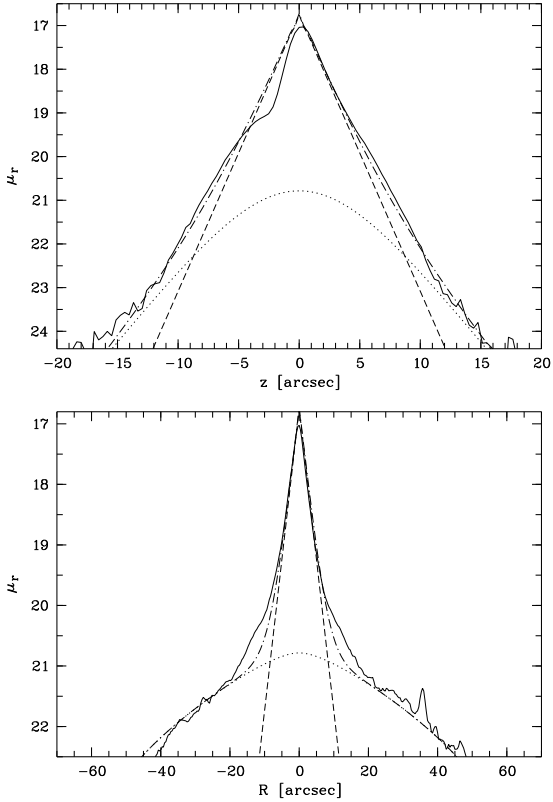,width=8cm,angle=0}
\caption{Minor and major axes profile for ESO575059r15E2 (solid lines) 
together with the best disk model (dotted), the best bulge model (dashed), 
and the resulting combined profile (dotted-dashed).}
\label{bulge}
\end{figure}

We have studied the influence of the bulge for some of our objects including 
the earliest type galaxy in our sample (ESO 575-059) presented here.
We have subtracted our derived disk model from the galaxy and then tried to 
find the best representation for the remaining bulge by a de Vaucouleurs 
$r^{1/4}$ or an exponential model. 
Taking the slope of the vertical profile at $R\!=\!0$ and a fixed axis ratio, 
we have constructed the 2-dimensional model of the bulge. 
In agreement with Andredakis et al. (\cite{andre}) we find, that bulges of 
early type galaxies are better fitted by an exponential profile than by a 
$r^{1/4}$.
Figure \ref{bulge} shows the resulting vertical and radial cuts for 
ESO 575-059 together with the models.   
Despite the deviation between $R\!=\!10''$ and $R\!=\!20''$  which could be 
attributed to an additional component (inner disk or bar), we do not find any 
evidence for changing our disk model due to the influence of the bulge.  \\  
Therefore we conclude, that it is possible to nearly avoid any influence of 
the central component by fitting outside the clearly visible bulge region.
\subsubsection{The influence of dust}
\label{dust}
Dust disturbs the light profile by a combination of absorption and extinction 
and the net effect has to be calculated by radiation transfer models. 
Therefore it is not obvious that outside the ``visible'' dust lane, which is 
excluded for the fitting area as a first step, the dust will not play a major 
role in shaping the light distribution.
Xilouris et al. (\cite{xil99}), Bianchi et al. (\cite{simone}), 
de Jong (\cite{dj1}), and Byun et al. (\cite{byun}) have recently addressed 
this problem in more detail.
Although they investigated the influence of the dust on the light distribution
by quoting best fit structural parameters for the star-disk as well as the 
dust-disk, they did not quantify the influence on the star-disk parameters 
derived by standard fitting methods without dust.
Even Kylafis \& Bahcall (\cite{kylafis}) state within their fundamental paper 
on finding the dust distribution for NGC 891 that ``in order to avoid 
duplication of previous work, we will take ... (the values for the 
star-distribution estimated by standard fitting methods)''.  \\   
We checked the influence of the dust on our determined parameter 
set, by studying simulated galaxy images with three different dust 
distributions.
These {\it dusty galaxies} were kindly provided by Simone Bianchi 
who calculated images with known input parameters for the star and dust 
distribution with his Monte Carlo radiative transfer method (Bianchi et al. 
\cite{simone}). 
We have defined a {\it worst}, {\it best}, and {\it transparent} case, according to 
the dust distributions presented by Xilouris et al. (\cite{xil99}), of our 
mean stellar disk ($R_{\rm co}\!=\!2.9$, $h/z\!=\!4$, and $f_1(z)$). 
The {\it worst} case is calculated with $\tau_R=0.51$, 
$h_{\rm d}/h_{\rm *}=1.55$, 
and $z_{\rm d}/z_{\rm *}=0.75$, the {\it best} case with $\tau_{\rm R}=0.20$, 
$h_{\rm d}/h_{\rm *}=1.08$, and $z_{\rm d}/z_{\rm *}=0.32$, and a {\it transparent} case 
without dust. 
To be comparable we used the same method for selecting the fitting region by 
masking the ``visible'' dust lane and reserving a typical area for a possible 
bulge component using mean values for the {\it transparent} case.
In contrast to our standard procedure we do not restrict the inclination
range from the appearance of the dust lane, but specify the best $SQ$ model 
in the range $i=80\degr\!-\!90\degr$ in Table \ref{dustvgl}. For the models 
marked with a '$\star$' we pretend the correct input inclination. 
Table \ref{dustvgl} demonstrates, that even for the worst case we are able to 
reproduce the input parameters within the range of the typical 20\% error 
discussed in Section \ref{comp}. It should be mentioned that for each case
we overestimate the input scalelength and -height, whereas the determination 
of $R_{\rm co}$ does not depend on the dust distribution. The implication on the
distribution of the ratio $R_{\rm co}/h$ will be discussed in a forthcoming 
paper.   
\begin{table}[h]
 \caption[]{Comparison of the results for the stellar disk 
(input: $z_{\rm *}\!=\!8.4$, $h_{\rm *}\!=\!33.7$, and $f(z)\!=$exp) 
from our fitting procedure for the three different dust distributions.}
 \label{dustvgl}
\begin{tabular}{l c |  c c c c r r}
\hline
case & $i_{\rm in}$ &$f_n$ & $i_{\rm out}$ & $z_{\rm *}$ & $h_{\rm *}$
&$\Delta z_{\rm *}$& $\Delta h_{\rm *}$   \\
&\raisebox{+0.2ex}[1.0ex]{[$\;\degr\;$]}
&
&\raisebox{+0.2ex}[1.0ex]{[$\;\degr\;$]}
&\raisebox{+0.2ex}[1.0ex]{ [$\;\arcsec\;$] }
&\raisebox{+0.2ex}[1.0ex]{ [$\;\arcsec\;$] } 
&\raisebox{+0.2ex}[1.0ex]{ [\%] }
&\raisebox{+0.2ex}[1.0ex]{ [\%] } \\
\hline\hline 
{\it trans.}  &87.5        &1&87.5 &8.5 &34.0 &$1.2$ &$0.9$\\
{\it best }   &85.0        &1&84.5 &8.5 &35.0 &$1.2$ &$3.9$\\
{\it best }   &87.5        &1&86.5 &8.6 &35.5 &$2.4$ &$5.3$\\
{\it best }   &90.0        &1&87.0 &8.7 &36.0 &$3.6$ &$6.8$\\
{\it best}$\star$    &90.0 &1&90.0 &8.8 &35.0 &$4.8$ &$3.9$\\
{\it worst}   &85.0        &2&85.0 &8.9 &39.9 &$6.0$ &$18.4$\\
{\it worst }  &87.5        &2&86.0 &8.8 &39.1 &$4.8$ &$16.0$\\
{\it worst}   &90.0        &2&88.0 &9.7 &39.4 &$15.5$&$16.9$\\
{\it worst}$\star$   &90.0 &2&90.0 &9.8 &39.1 &$16.7$&$16.0$\\
\hline
\end{tabular}
\end{table}
\subsection{Comments on individual galaxies}
Trying to adapt a simple, perfect, and exact symmetric model to real galaxies
always implies a compromise between the degree of any deviation and the
final model (Section \ref{pec}). 
The following list will provide some typical caveats found
during the fit procedure which will characterise the quality of the 
specified model for individual galaxies.\\
{\bf ESO 112-004}: warped, asymmetric, central part slightly tilted compared to
disk, after fitting still remaining residuals \\
{\bf ESO 150-014}: slightly warped, minor flatfield problems \\
{\bf NGC 585}: remaining residuals \\  
{\bf ESO 244-048}: possible two component system, slope of inner radial 
profile significantly higher than of an outer one, final model fits the inner 
parts\\
{\bf NGC 973}: one side disturbed by stray light of nearby star, seems to be 
radially asymmetric, remaining residuals \\
{\bf UGC 3425}: superimposed star on one edge \\
{\bf NGC 2424}: model does not fit very well without obvious reason \\    
{\bf ESO 436-034}: strong bulge component, possibly barred, hard to pinpoint 
final model, remaining residuals \\  
{\bf ESO 319-026}: outer parts show u-shaped behaviour, remaining residuals, 
therefore large ($\pm 2 \degr$) difference in inclination angle \\
{\bf ESO 321-010}: u-shaped, no clear major axis visible, therefore uncertain 
rotation angle, bar visible, bulge rotated against disk \\  
{\bf NGC 4835A}: strong residuals \\
{\bf ESO 446-018}: the different sides of the disk are asymmetric 
visible in radial profiles and on the contour plot \\
{\bf IC  4393}: similar to NGC 4835A \\    
{\bf ESO 581-006}: galaxy shows typical late type profile, $R_{\rm co}$ 
questionable, but nevertheless final model seems to fit well \\  
{\bf ESO 583-008}: disturbed by superimposed star, shows warp feature
and a bar structure, $R_{\rm co}$ questionable,  remaining residuals\\
{\bf UGC 10535}: one side slightly extended \\  
{\bf NGC 6722}: only one side observed, bulge rotated against disk, barred, 
strongly disturbed by dust absorption, radial extension visible, therefore 
$R_{\rm co}$ should be treated with caution\\  
{\bf ESO 461-006}: minor flatfield problem seems to cause asymmetry, 
although model looks fine \\  
{\bf IC  4937}: similar to NGC 6722, dominating bulge, small disk,
model significantly different compared to the i and r image, model possibly
hampered by strong dust lane\\
{\bf ESO 578-025}: bar visible \\  
{\bf ESO 466-001}: maybe two components, final model represents only inner 
part, outer part clearly different from normal disk component \\
{\bf ESO 189-012}: slightly warped \\ 
{\bf ESO 533-004}: similar to NGC 4835A, model fits the whole galaxy, leaving
more or less no bulge component \\
{\bf IC  5199}: slightly radial asymmetric\\ 
{\bf ESO 604-006}: only one side observed, bar structure visible   
\subsection{Comments on some rejected galaxies}
\label{pec}
The model limitations described above constrain the application of our fitting process. 
Therefore we had to exclude about 20 galaxies from our original sample. 
They all show significant deviations from the simple geometry and an 
inclusion of their parameters obtained by forcing the model to fit the data 
will spoil the resulting parameter distribution. \\
One larger group classified mainly as S0 galaxies
(e.g. NGC 2549, ESO 376-009, NGC 7332, ESO 383-085, ESO 506-033) shows a 
completely different behaviour of the luminosity distribution in the outer 
parts compared to the other galaxies. They all show an additional  
component, mainly characterised as an elliptical envelope. This is already 
visible in the contour plot, but becomes even more evident in a radial cut parallel to the 
major axis. In these cases the usual common curved decline of the profile 
(e.g. ESO 578-025) is missing, and is replaced by a more or less straight
decline into the noise level, sometimes even by an upwards curved profile.
Fitting these luminosity distribution by our one component exponential disk
with cut-off, will therefore naturally provide parameters qualitatively 
different compared to late type disks. This will be discussed in detail in a 
forthcoming paper. \\
Another group consists of galaxies dominated mainly by their bulges,
whereas the disk is only an underlying component, partly characterized as 
having thick boxy bulges (Dettmar \& L\"utticke \cite{db}), e.g. IC 4745, 
ESO 383-005, although there are also pure elliptical bulges 
(e.g. ESO 445-049, NGC 6948).\\
In the case of ESO 383-048 and ESO 510-074 the radial profiles clearly 
indicate that a more complex model will be needed to fit these kind
of multicomponent galaxies.   
Galaxies like UGC 7170 or ESO 113-006 were excluded due to their strong 
warps, which made it impossible to fit the model in a consistent way. 
Mainly late type galaxies such as ESO 385-008, IC4871, UGC 1281, or 
ESO 376-023 show a too patchy and asymmetric light distribution, that 
any attempt to fit the profiles will give only very crude, low quality  
parameters. UGC 11859 and UGC 12423 were rejected due to their
thin faint disks, which will maybe overcome by taking new images with longer 
integration times to get a higher signal-to-noise ratio, whereas NGC 5193A is
completely embedded into the surface brightness distribution of its near 
companion.
%
\begin{acknowledgements}
This work was supported by the \emph{Deut\-sche For\-schungs\-ge\-mein\-schaft, 
DFG\/}. This research has made use of the NASA/IPAC Extragalactic Database
(NED) which is operated by the Jet Propulsion Laboratory, California
Institute of Technology, under contract with the National Aeronautics and
Space Administration. We have made use of the LEDA database 
(www-obs.univ-lyon1.fr). The authors wish to thank Simone Bianchi,
who kindly provided us {\it dusty-galaxy images} produced with his radiative 
transfer code.     

\end{acknowledgements}

\newpage
\newpage
\appendix
\section{Contour plots and radial profiles}
\label{app}
The following figures show in the top panel selected cuts parallel to the major
axis (full line) as well as the best fit model (dashed line) according to 
Table \ref{modelle}. The bottom panel shows isophote maps of the surface
brightness for the galaxies, rotated to the major axis. The faintest plotted 
contour is defined by a 3$\sigma$ criterion of the background.  
Table \ref{plots} lists the name {\scriptsize{\it (1)}}, image 
{\scriptsize{\it (2)}}, and the vertical positions of the plotted profiles 
parallel to the major axes in arcsec {\scriptsize{\it (3)}}. 
The contours are plotted equally spaced with 0.5 mag$/\sq\arcsec$ starting 
from the limiting surface brightness $\mu_{{\rm lim.}}$ {\scriptsize{\it (4)}}.
   \begin{table}[h]
      \caption[]{Plotted parameters}
\begin{tabular}{l c c c}
\hline
galaxy&image&$z$     &$\mu_{{\rm lim.}}$ \\
      &     &[$\;\arcsec\;$]&[mag$/\sq\arcsec$] \\
\rule[-3mm]{0mm}{5mm}
{\scriptsize{\raisebox{-0.7ex}{\it (1)}}}
&{\scriptsize{\raisebox{-0.7ex}{\it (2)}}}
&{\scriptsize{\raisebox{-0.7ex}{\it (3)}}}
&{\scriptsize{\raisebox{-0.7ex}{\it (4)}}} \\
\hline\hline \\
NGC 585   &  R20L1   & 3,6,9,12   & 24.7\\
ESO 244-048 &  r15E3  & $-$3,$-$9      & 24.8\\
NGC 973   &  R10L1   & $-$2,$-$4,$-$6,$-$8& 23.8\\
UGC 3326  &  R30L1   & 3,6,9      & 24.8\\
UGC 3425  &  R30L1   & $-$1,$-$5,$-$9   & 24.2\\
NGC 2424  &  R15L1    & 1,3,5,7    & 24.5\\
IC  2207   &  R10L2   & 2,6,10     & 24.1\\
ESO 564-027 &  r30E2  & 2,4,6      & 25.2\\
ESO 319-026 &  i30E3  & 2,4,6      & 24.4\\
ESO 319-026 &  r30E2  & 2,4,6      & 24.7\\
ESO 575-059 &  r15E2  & 3,6,9      & 24.4\\
ESO 578-025 &  g30E2  & 2,4,6      & 25.2\\
ESO 578-025 &  i30E2  & 2,4,6      & 23.3\\
ESO 578-025 &  r30E2  & 2,4,6      & 24.5\\
ESO 446-018 &  r30E2  & 2,4,6,8    & 25.1\\
IC  4393   &  r30E2   & $-$3,$\pm$6,$\pm$9 & 25.1\\
ESO 583-008 &  r30E3  & 2,4,6      & 25.0\\ 
UGC 10535  &  r25E2   & 2,4,6      & 25.5\\
IC  4937   &  g20E1   & $-$2,$-$4,$-$6   & 25.1\\
IC  4937   &  i20E1   & $-$2,$-$4,$-$6   & 23.3\\
ESO 528-017 &  i30E3  & 2,4,6      & 24.2\\
ESO 466-001 &  i40E3  & 1,3,5      & 23.8\\
ESO 189-012 &  i20E3  & $-$1,$-$3,$-$5   & 24.1\\
ESO 533-004 &  r20E1  & $-$3,$-$6,$-$9   & 24.6\\
IC  5199   &  i30E3   & $\pm$2,$\pm$4,$\pm$6& 23.8\\
\hline
\multicolumn{4}{l}{{\scriptsize remaining images already published in Paper I}}\\
\end{tabular}
         \label{plots}
  \end{table}
\clearpage
   \begin{figure*}
 \vspace*{1cm}
\psfig{figure=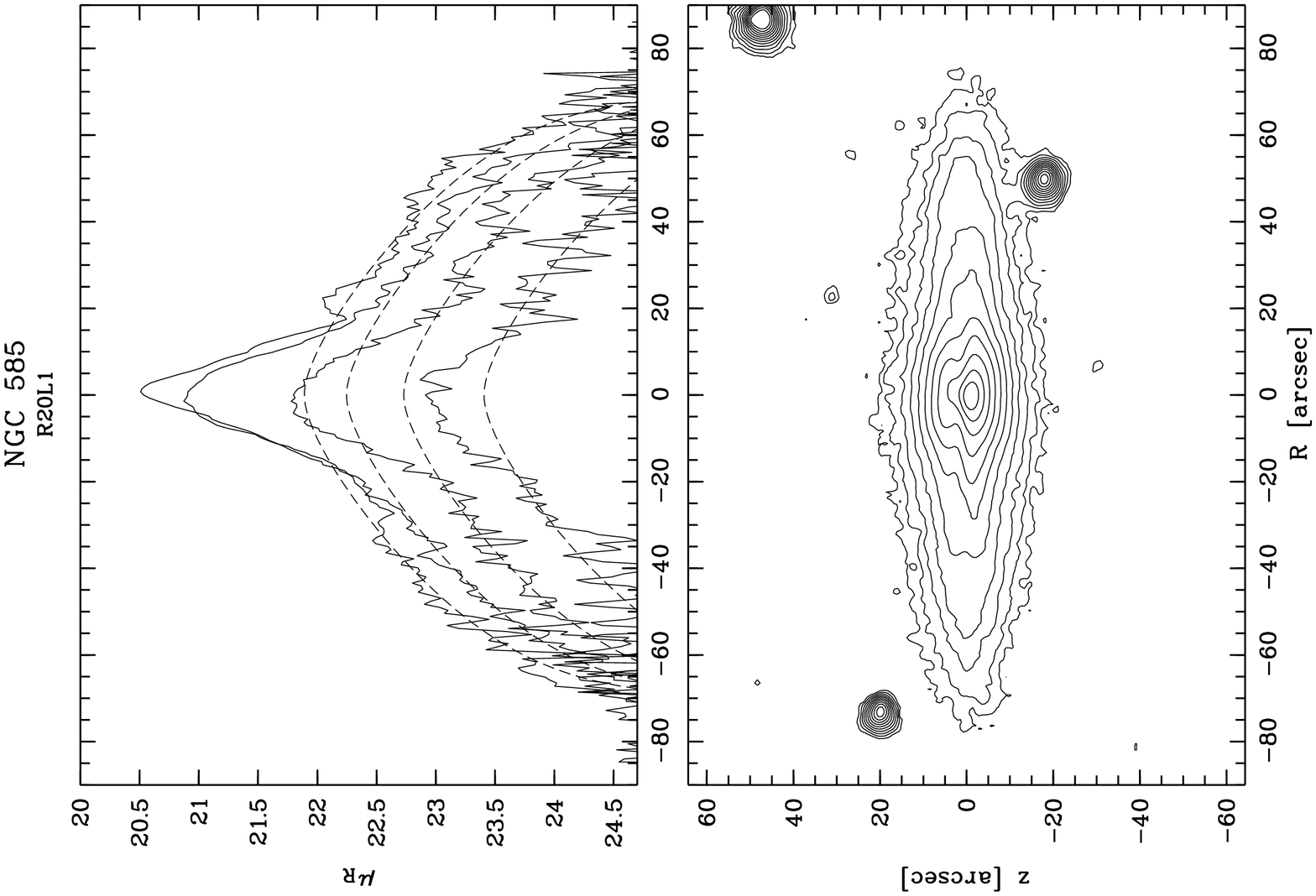,width=16cm,angle=90}
   \end{figure*}
   \begin{figure*}
\psfig{figure=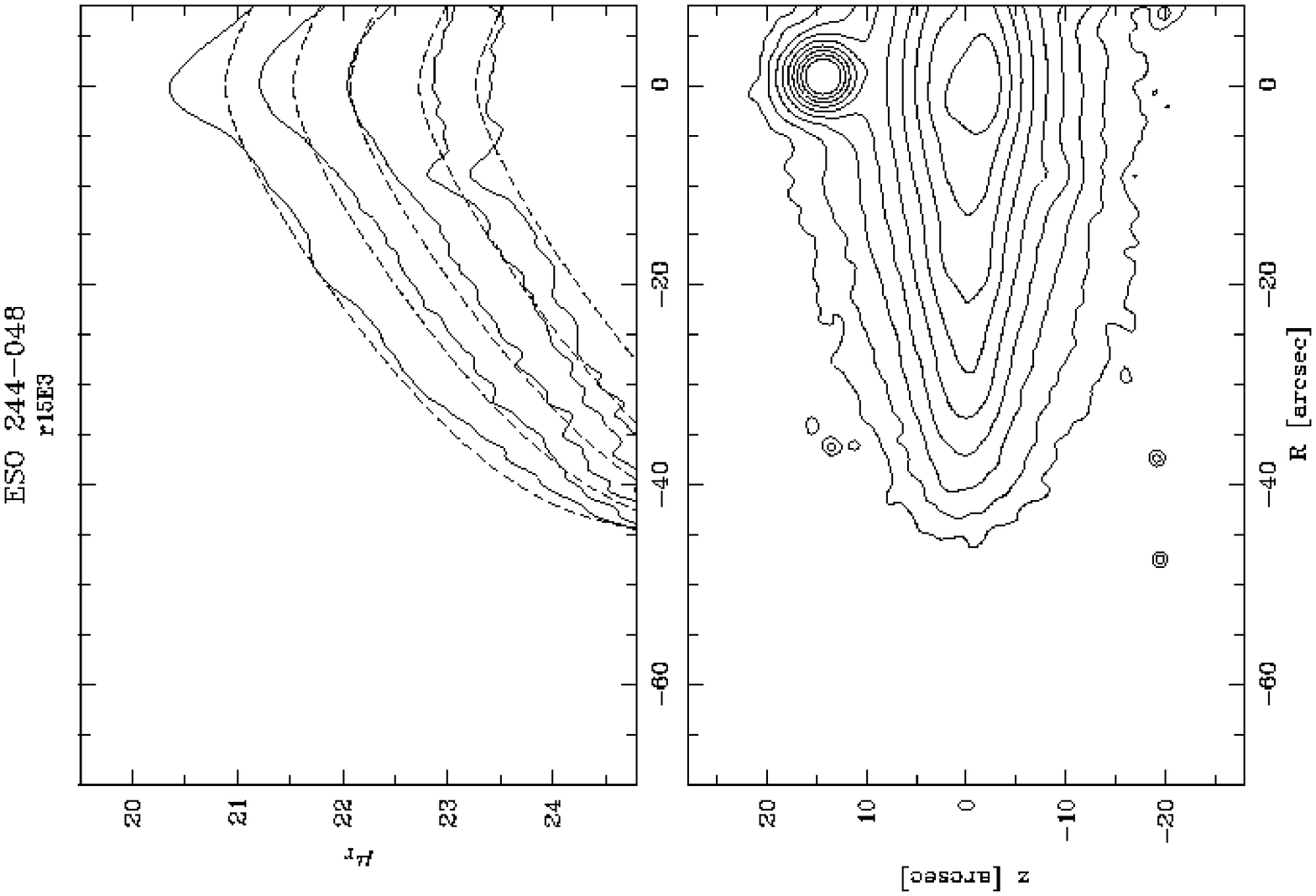,width=15cm,angle=90}
\end{figure*}
   \begin{figure*}
\vspace*{1cm}
\psfig{figure=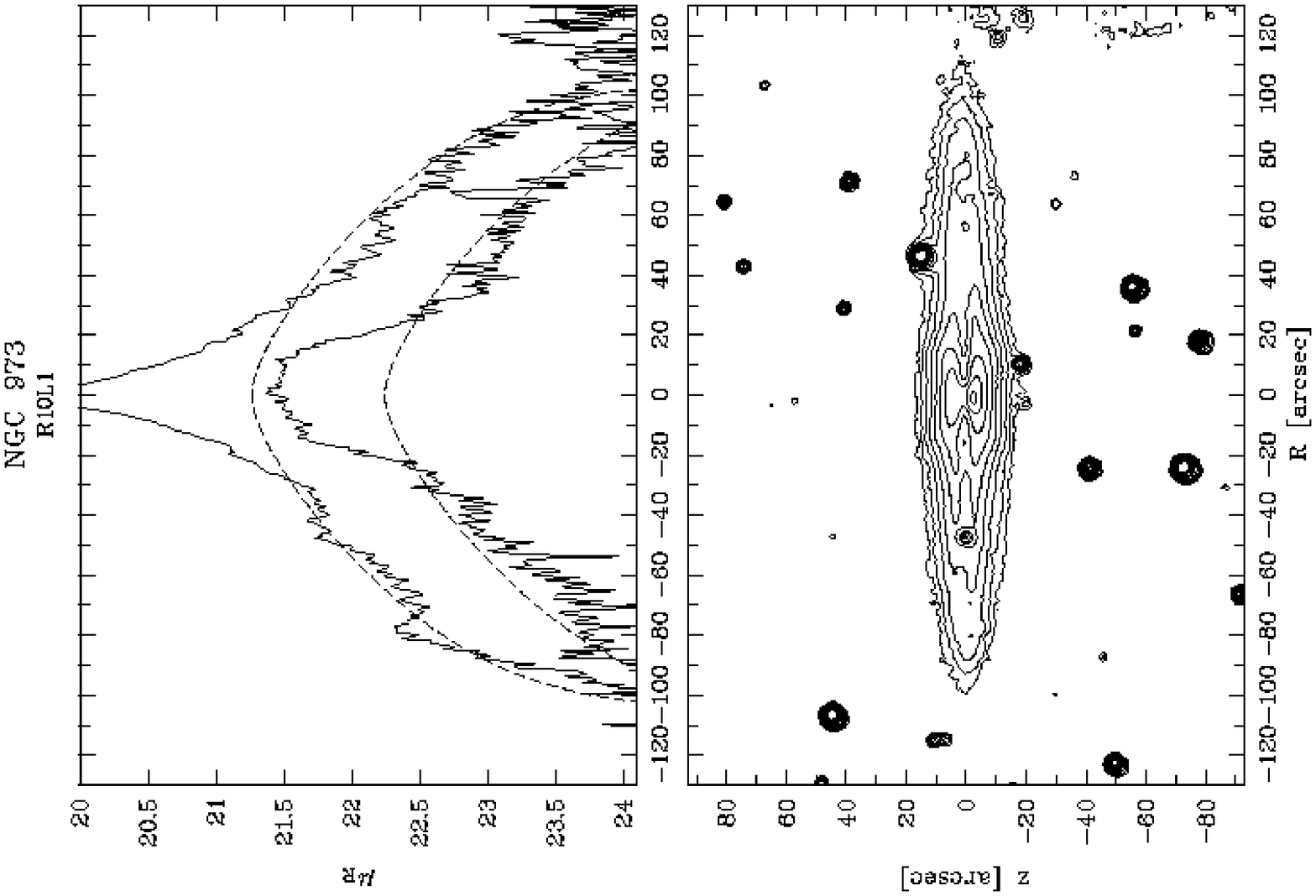,width=15cm,angle=90}
   \end{figure*}
   \begin{figure*}
\psfig{figure=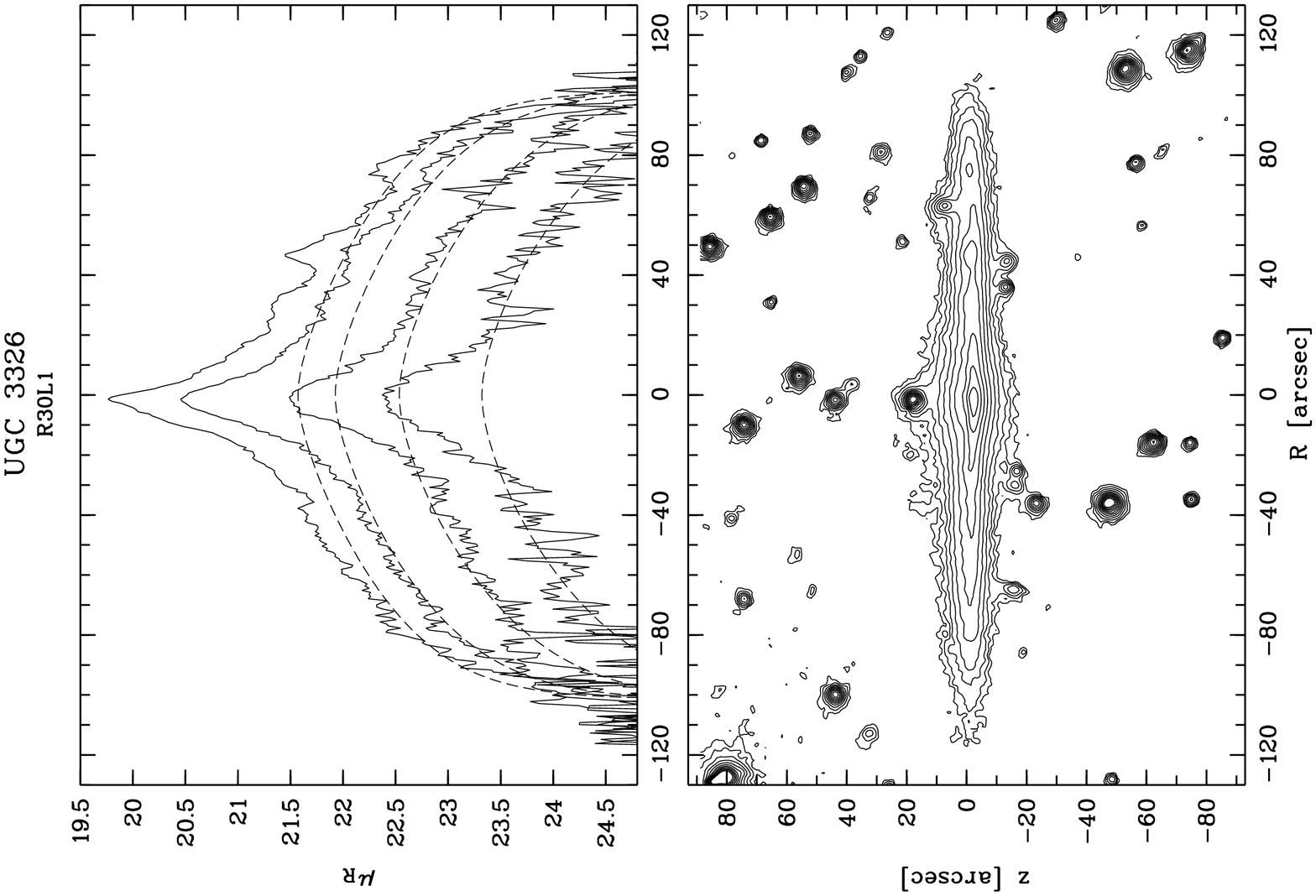,width=15cm,angle=90}
   \end{figure*}
   \begin{figure*}
\vspace*{1cm}
\psfig{figure=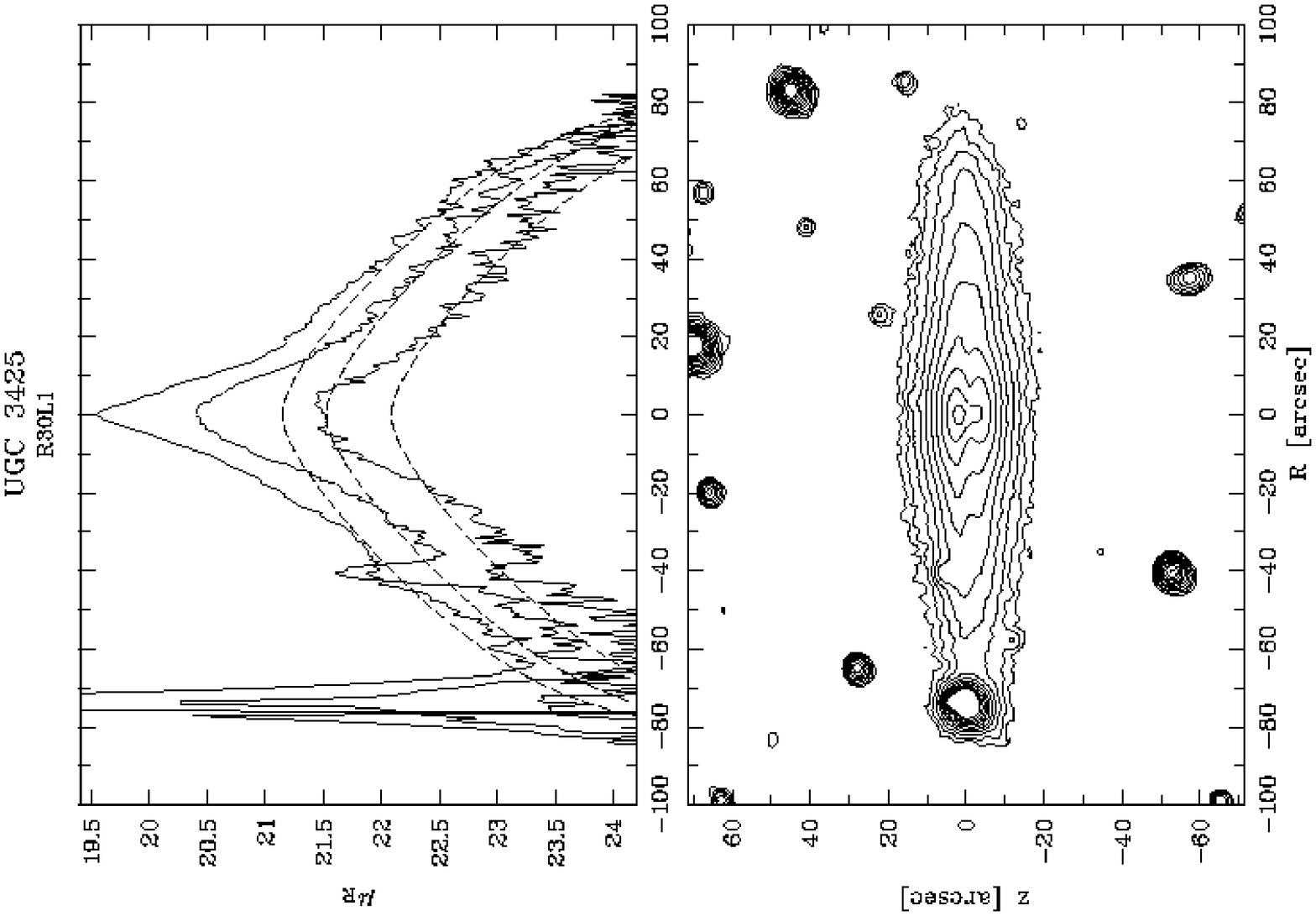,width=15cm,angle=90}
   \end{figure*}
   \begin{figure*}
\psfig{figure=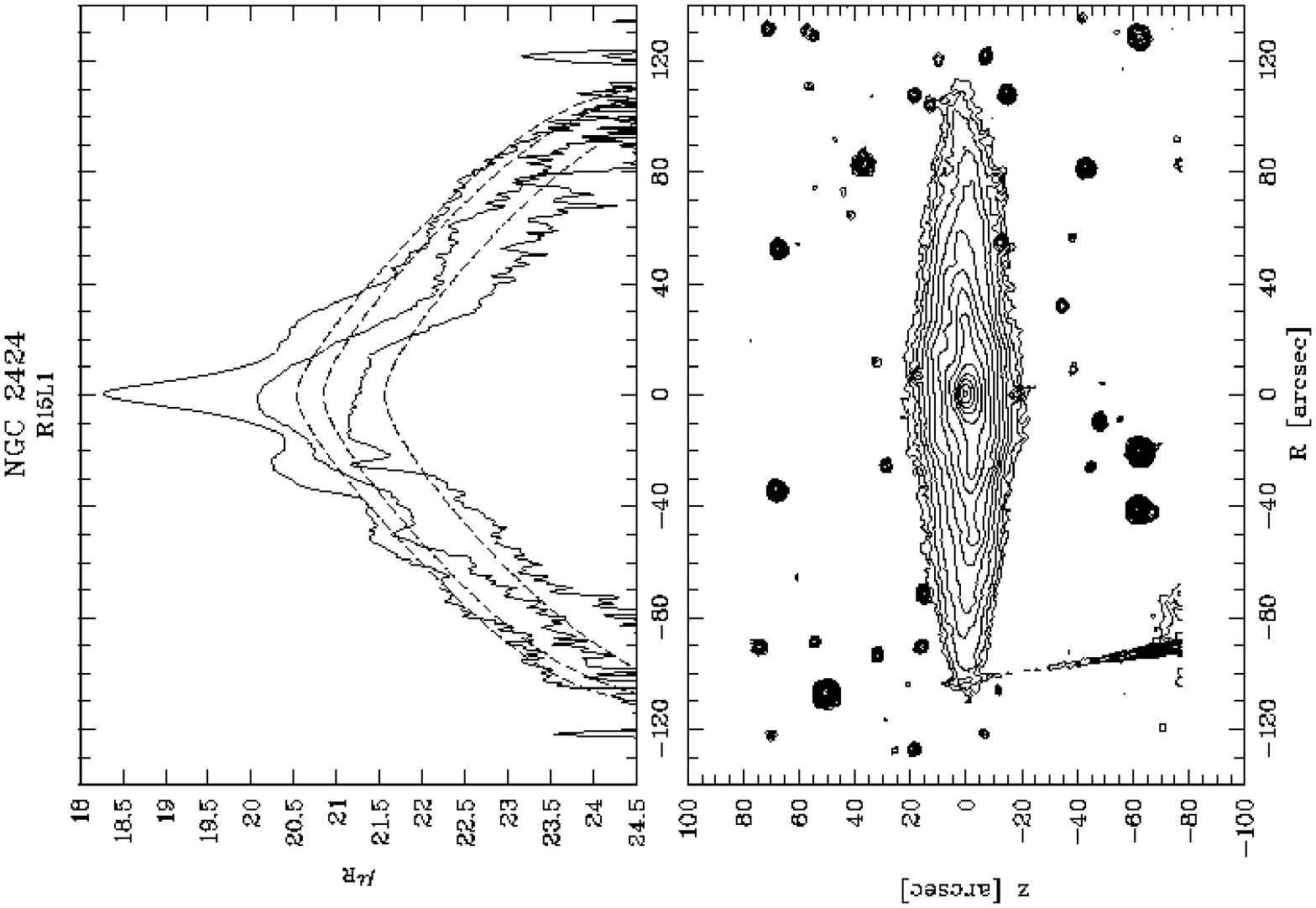,width=15cm,angle=90}
   \end{figure*}
   \begin{figure*}
\vspace*{1cm}
\psfig{figure=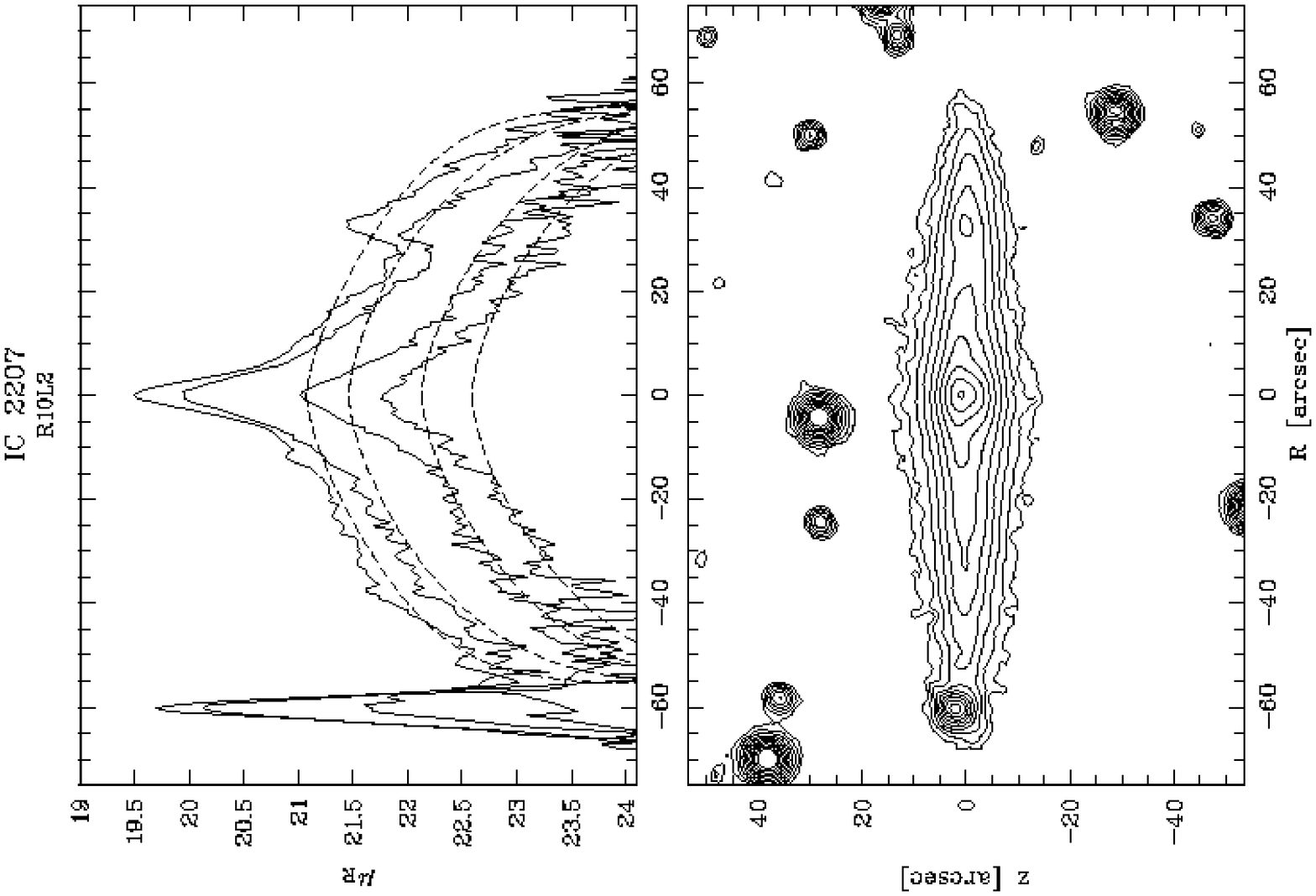,width=15cm,angle=90}
   \end{figure*}
   \begin{figure*}
\psfig{figure=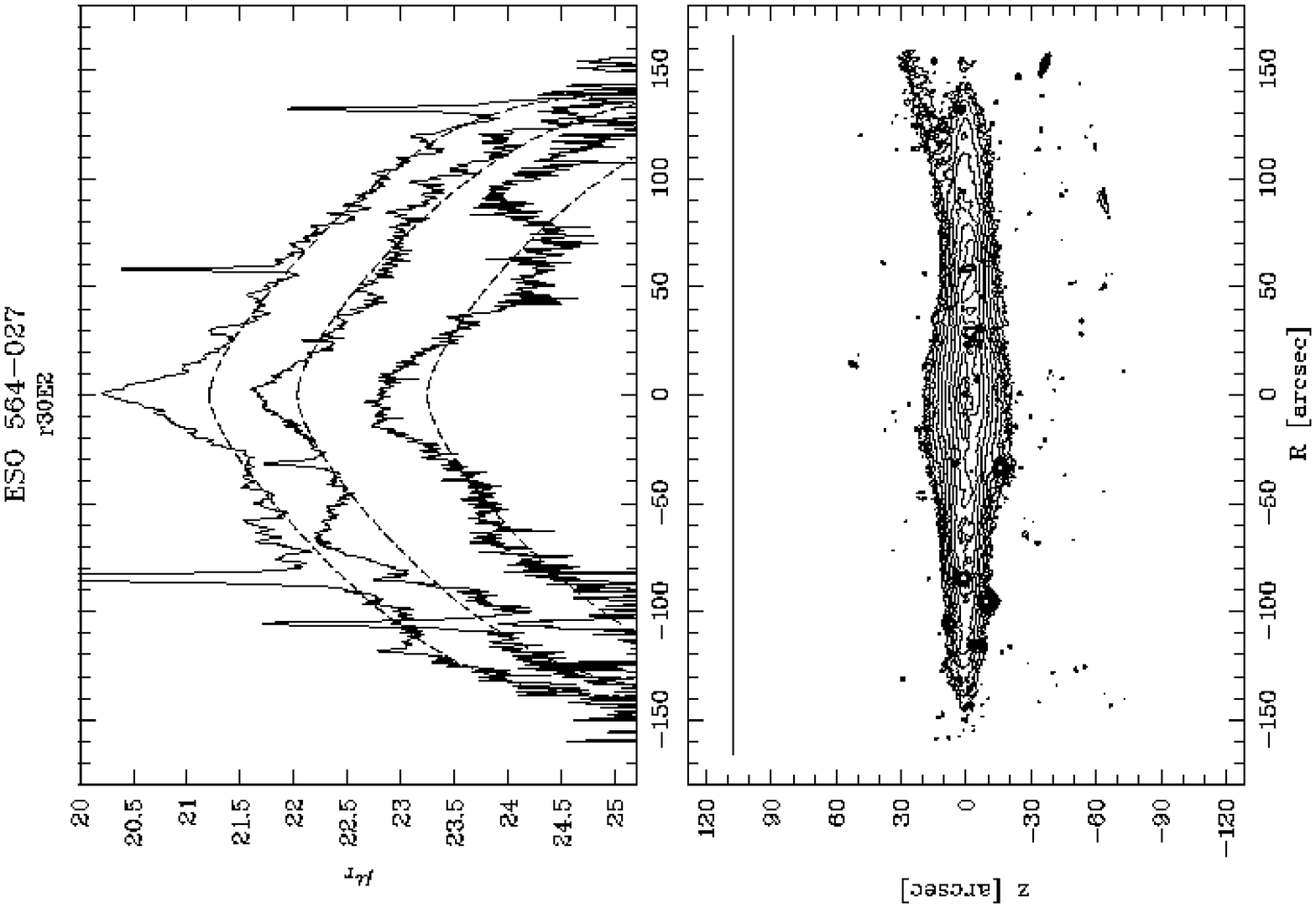,width=15cm,angle=90}
   \end{figure*}
   \begin{figure*}
\vspace*{1cm}
\psfig{figure=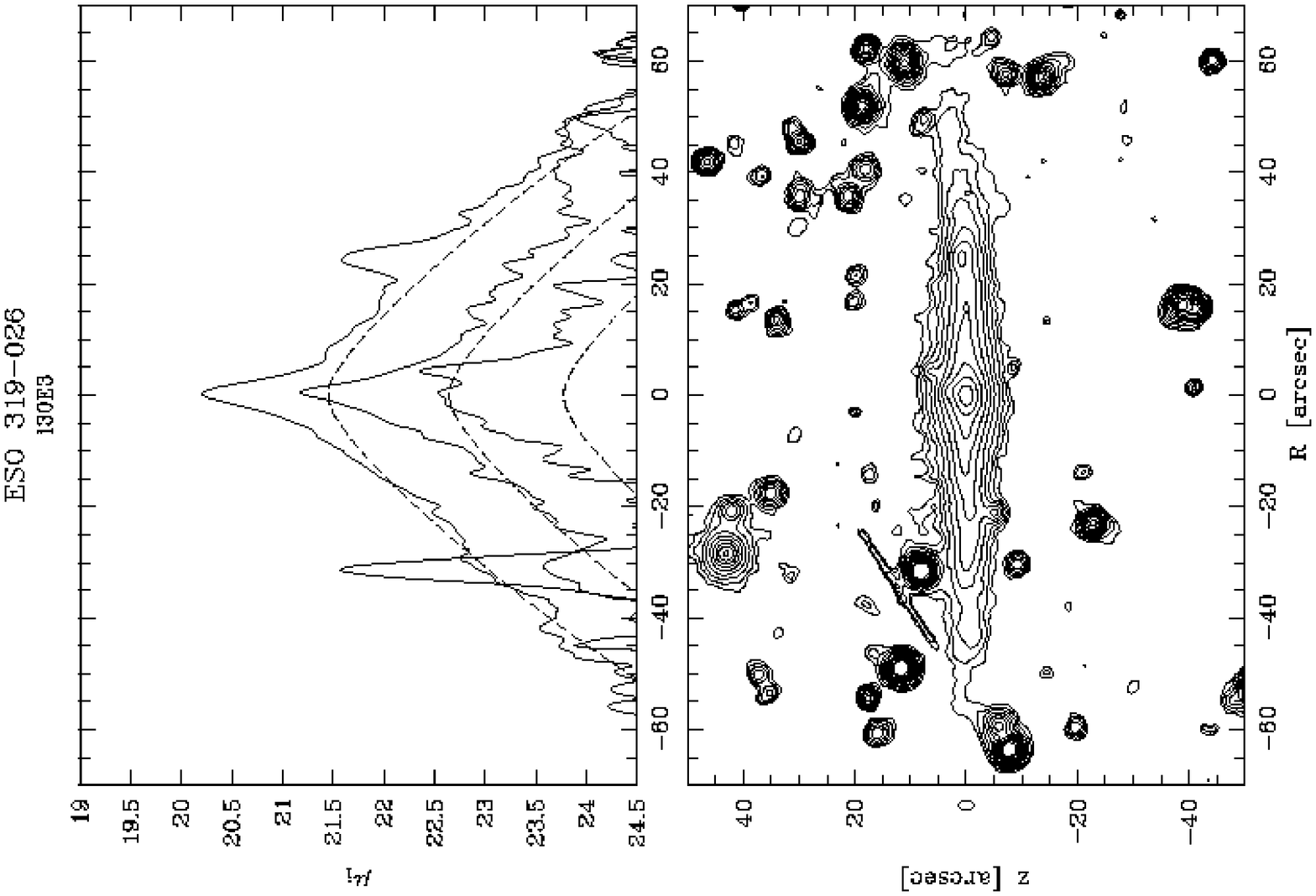,width=15cm,angle=90}
   \end{figure*}
\begin{figure*}
\psfig{figure=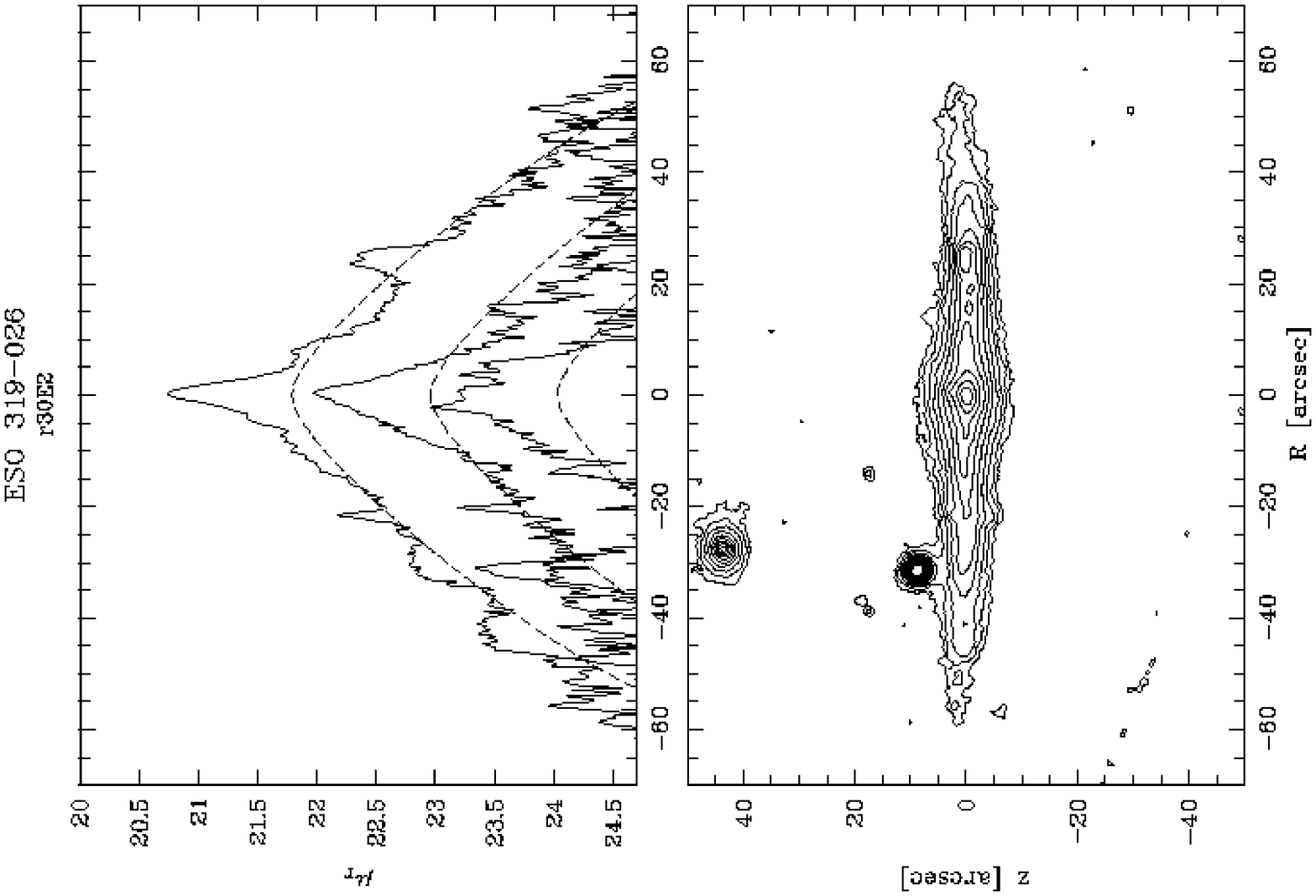,width=15cm,angle=90}
   \end{figure*}
\clearpage
\begin{figure*}
\vspace*{1cm}
\psfig{figure=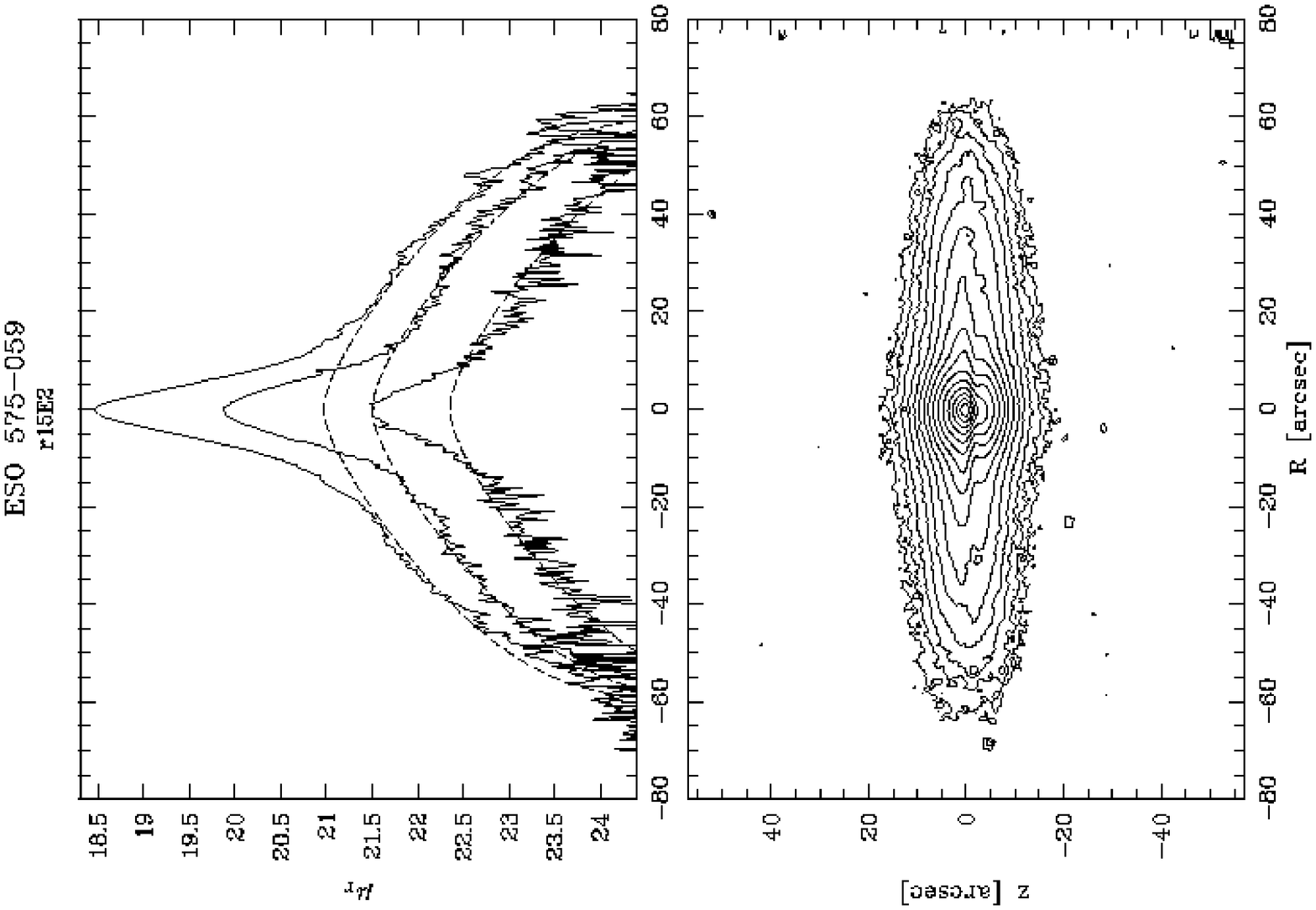,width=15cm,angle=90}
   \end{figure*}
\begin{figure*}
\psfig{figure=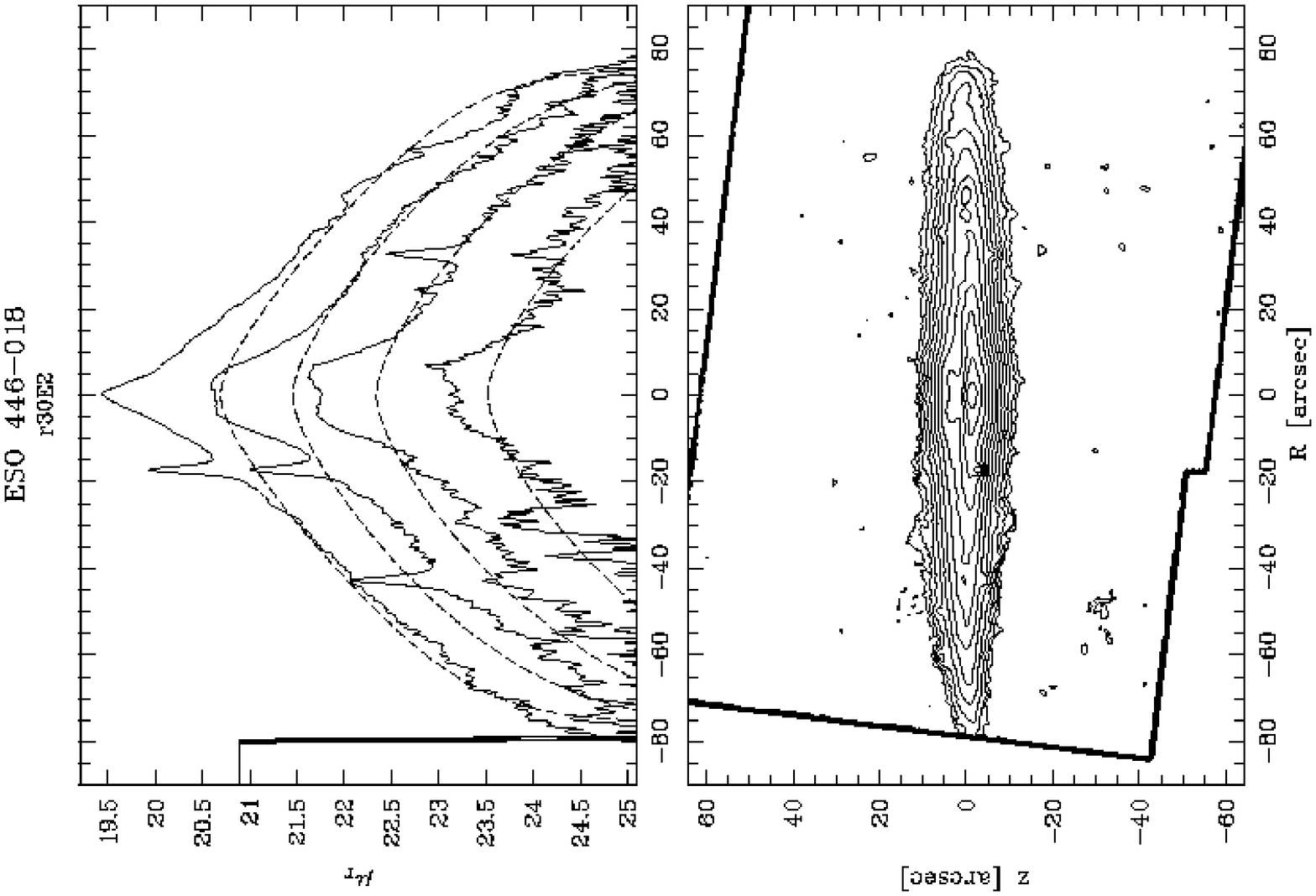,width=15cm,angle=90}
   \end{figure*}
\begin{figure*}
\vspace*{1cm}
\psfig{figure=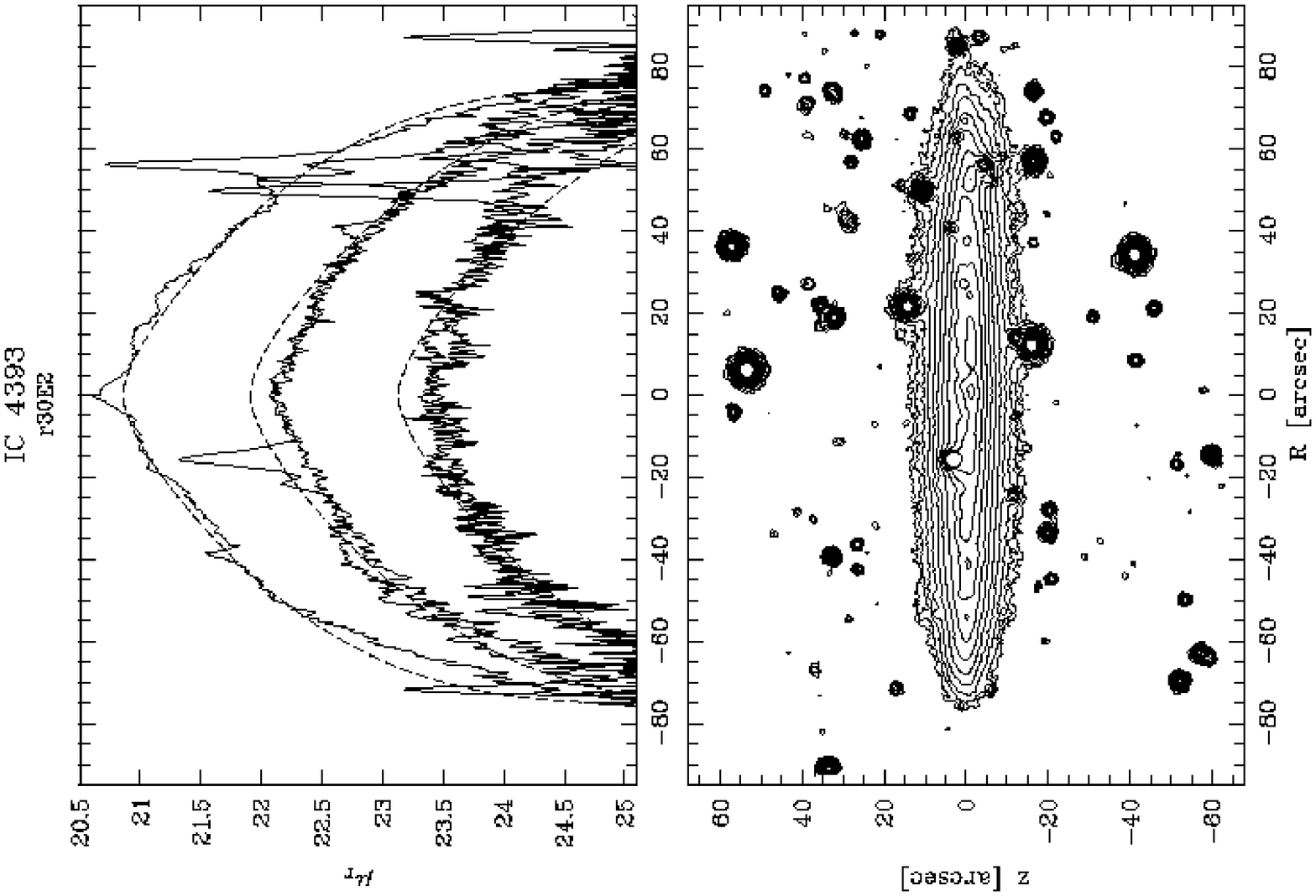,width=15cm,angle=90}
   \end{figure*}
\begin{figure*}
\psfig{figure=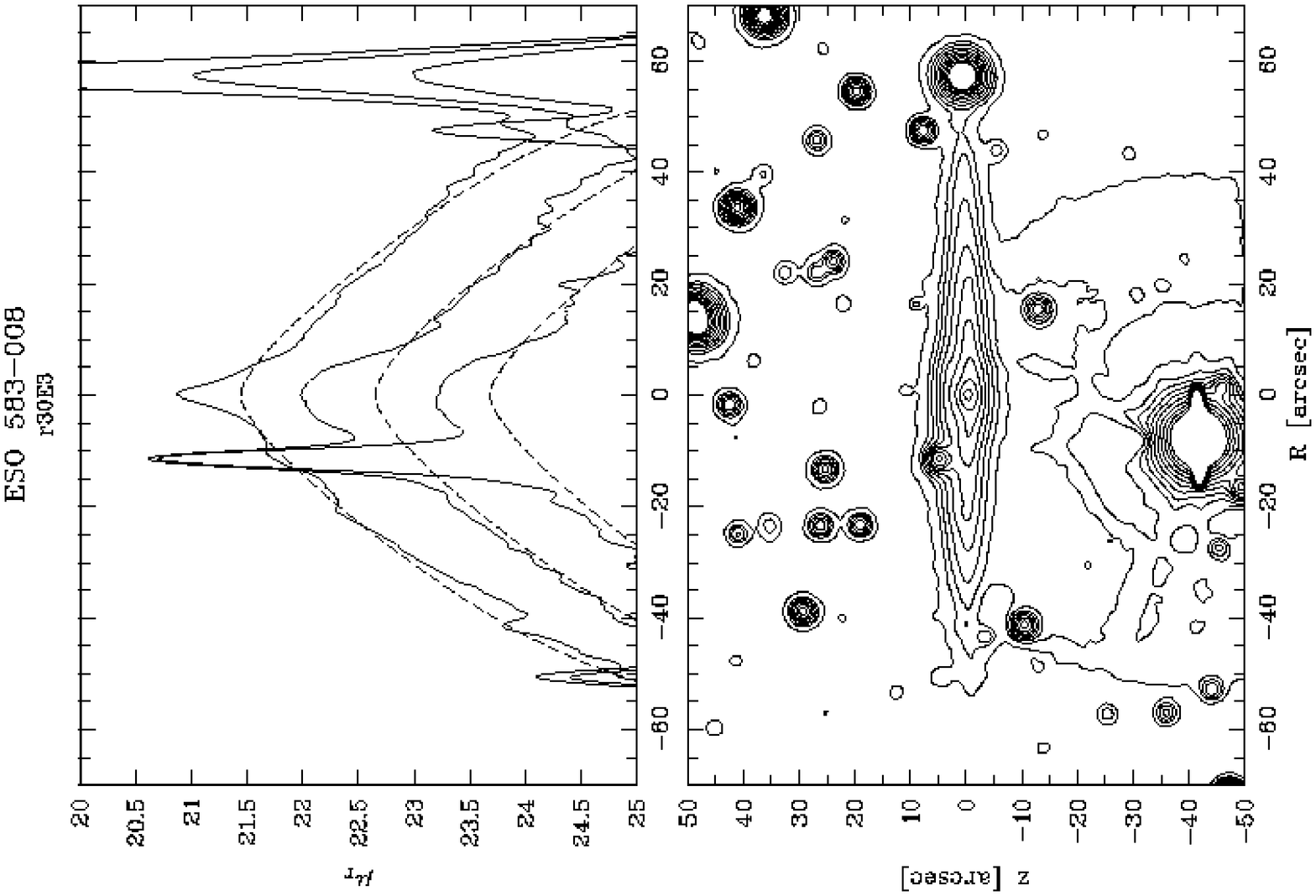,width=15cm,angle=90}
   \end{figure*}
\begin{figure*}
\vspace*{1cm}
\psfig{figure=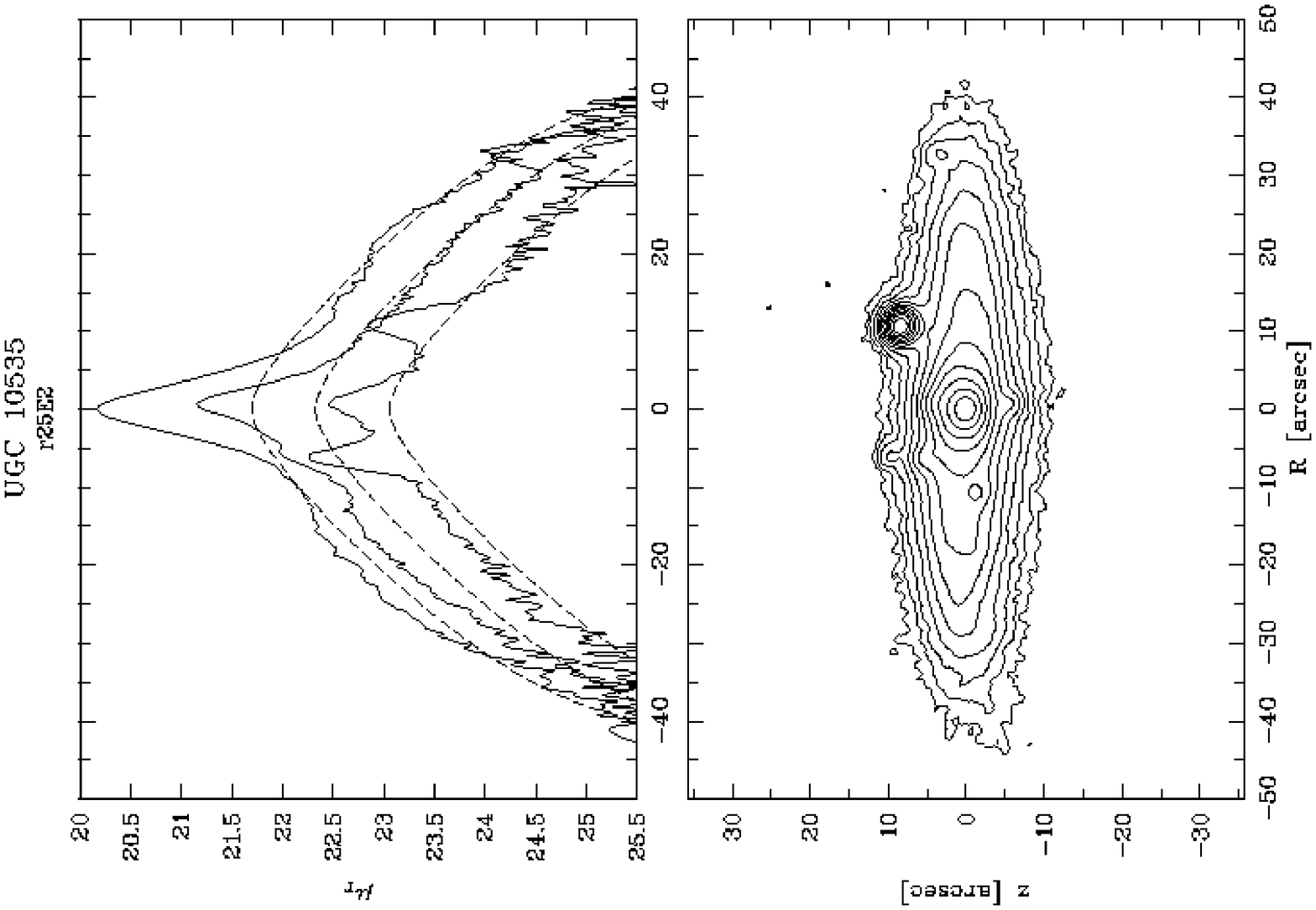,width=15cm,angle=90}
   \end{figure*}
\begin{figure*}
\psfig{figure=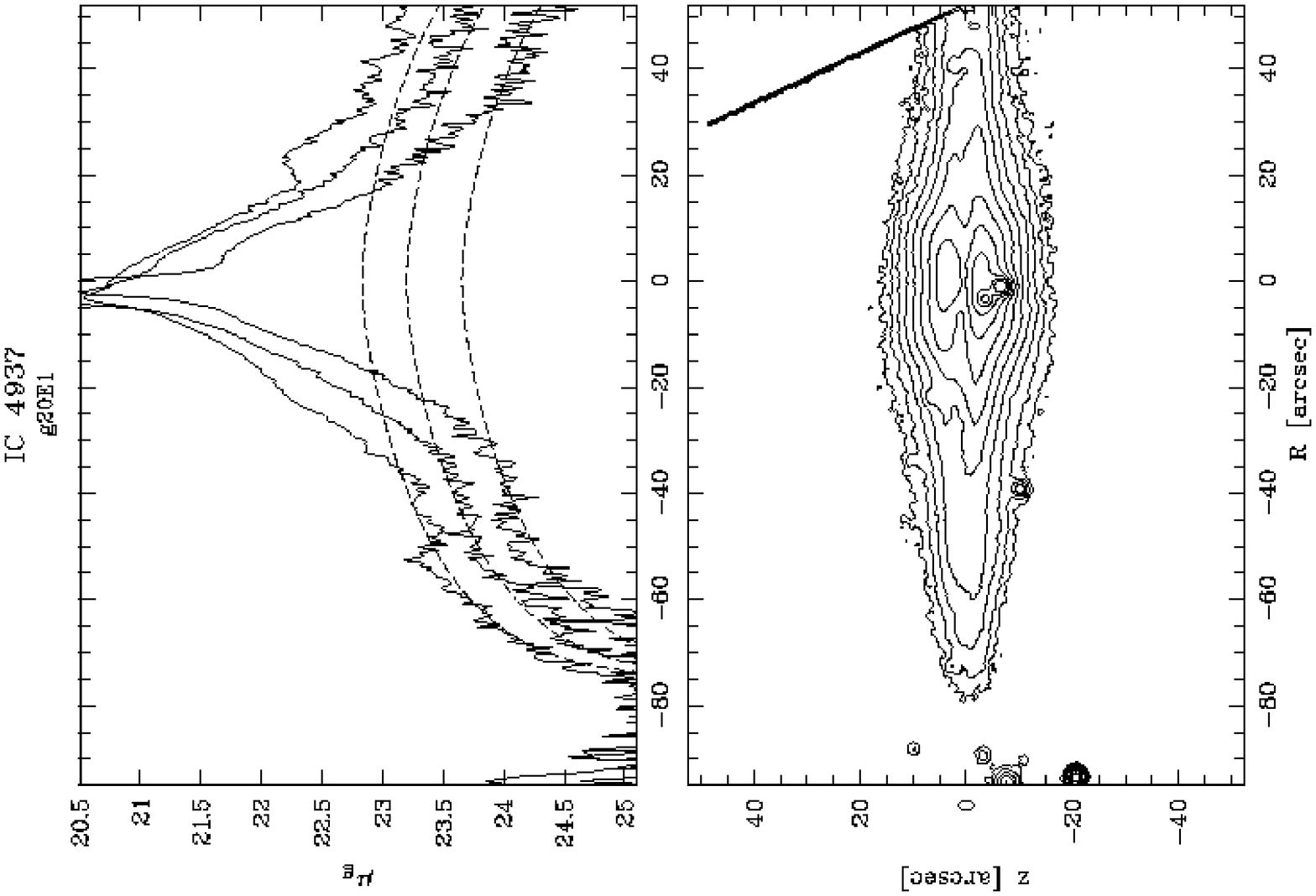,width=15cm,angle=90}
   \end{figure*}
\begin{figure*}
\vspace*{1cm}
\psfig{figure=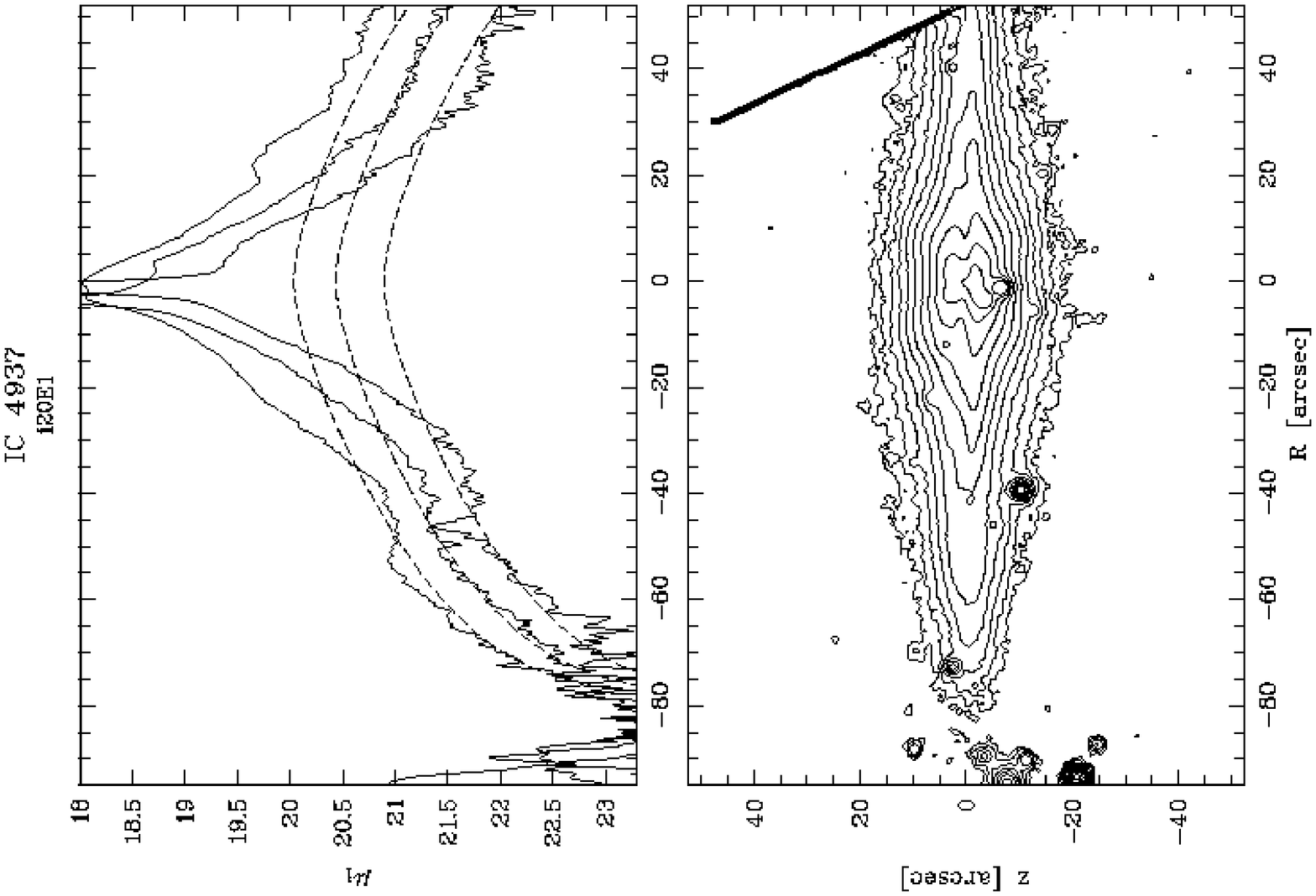,width=15cm,angle=90}
   \end{figure*}
\begin{figure*}
\psfig{figure=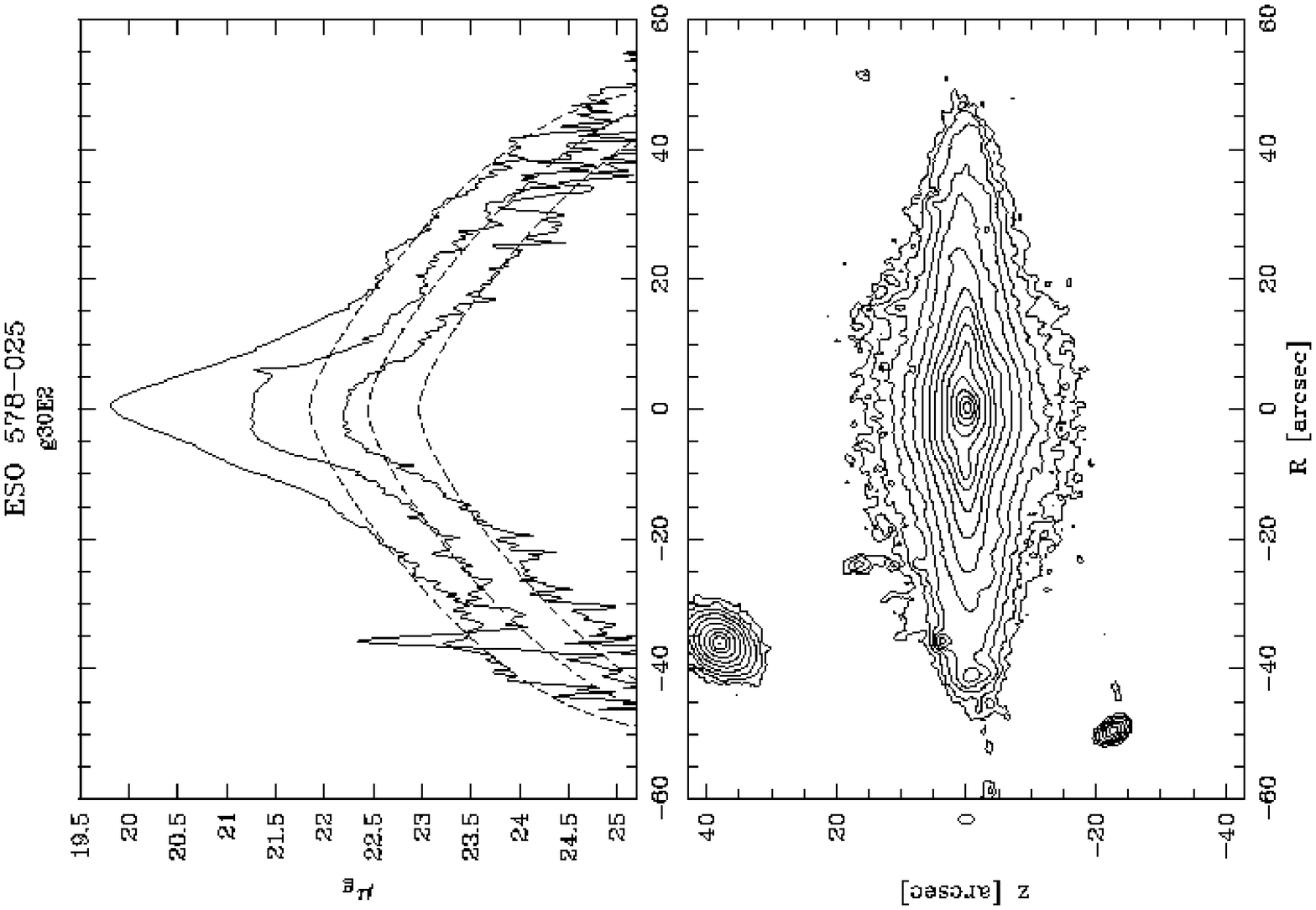,width=15cm,angle=90}
   \end{figure*}
\begin{figure*}
\vspace*{1cm}
\psfig{figure=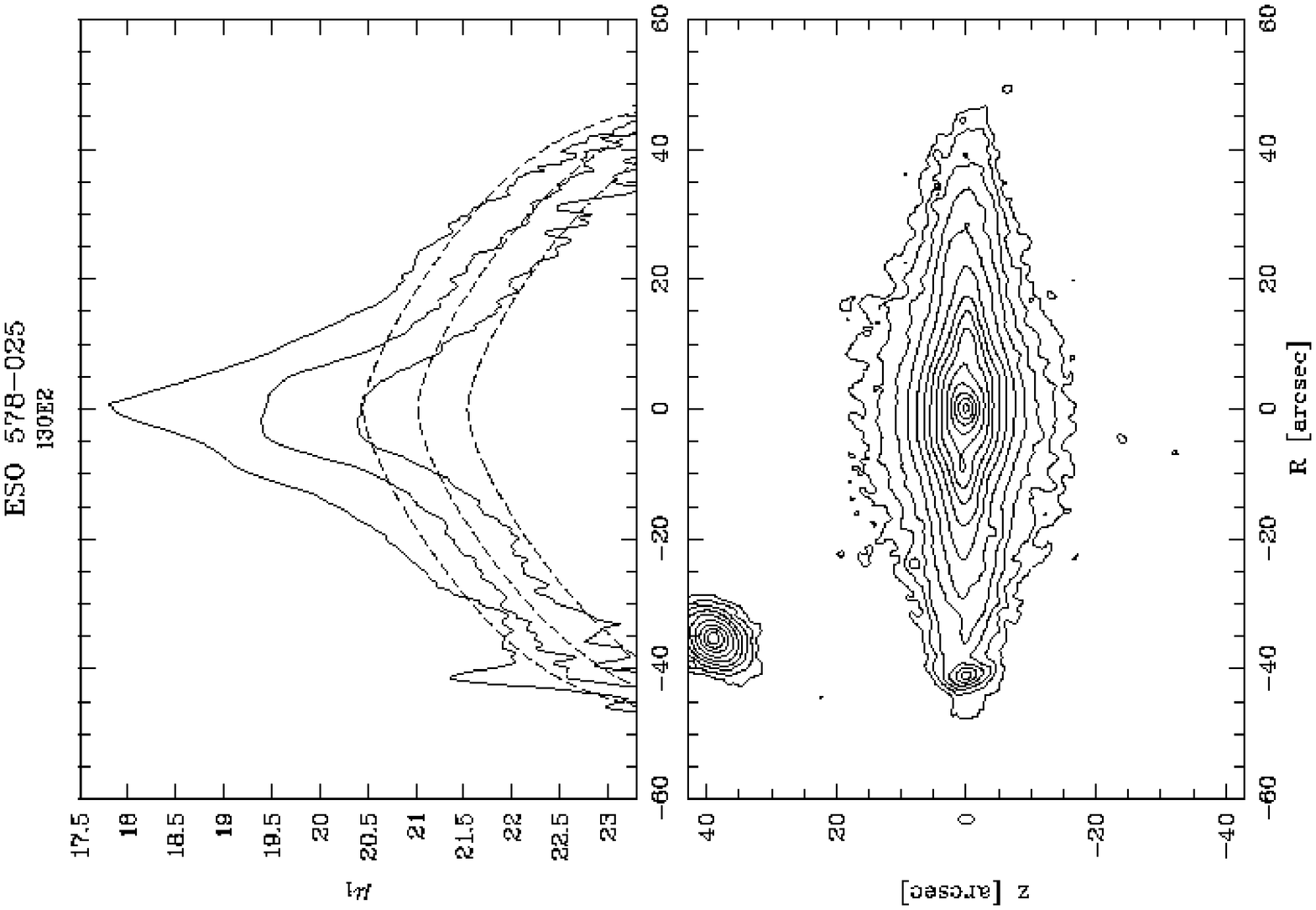,width=15cm,angle=90}
   \end{figure*}
\begin{figure*}
\psfig{figure=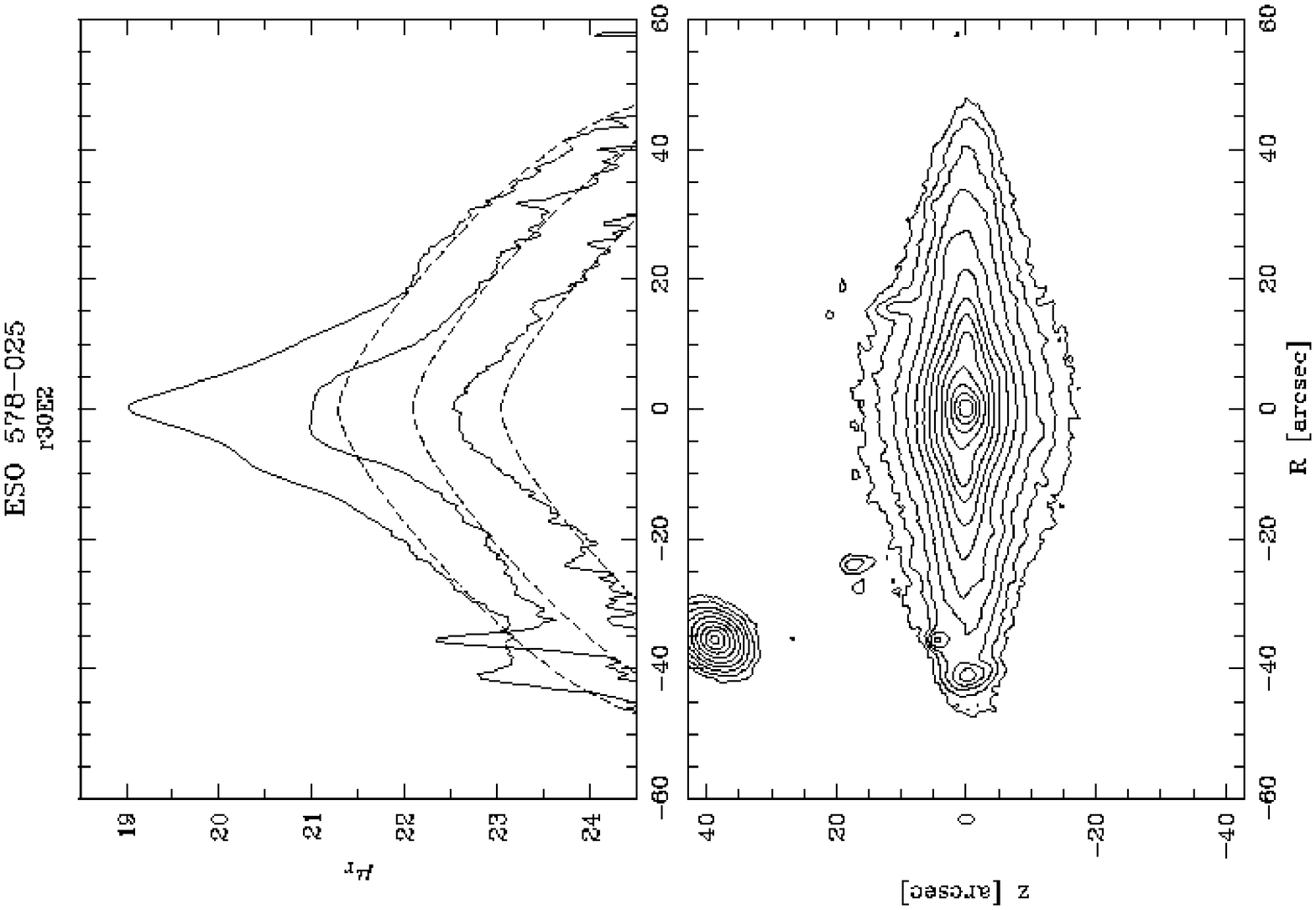,width=15cm,angle=90}
   \end{figure*}
\begin{figure*}
\vspace*{1cm}
\psfig{figure=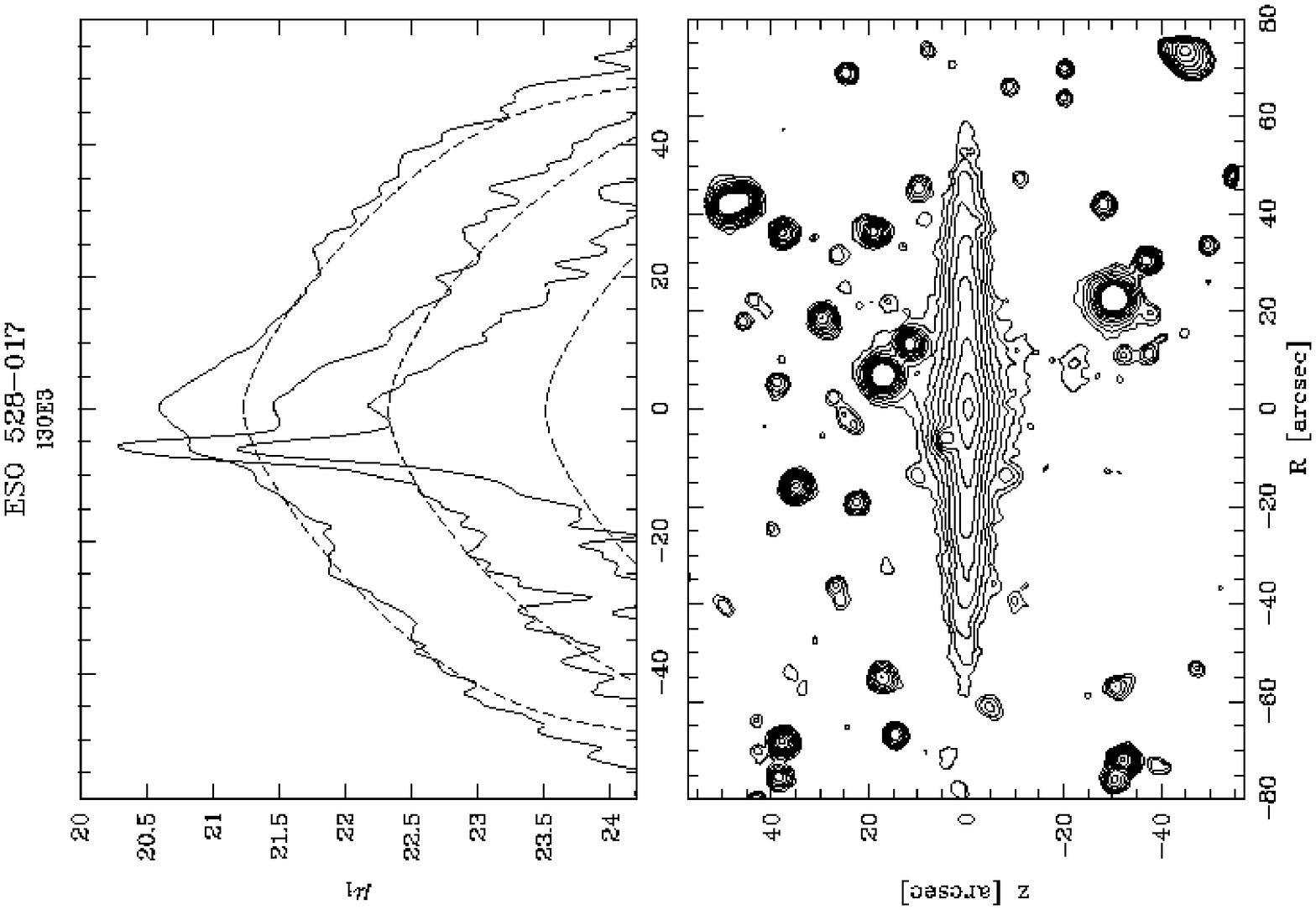,width=15cm,angle=90}
   \end{figure*}
\begin{figure*}
\psfig{figure=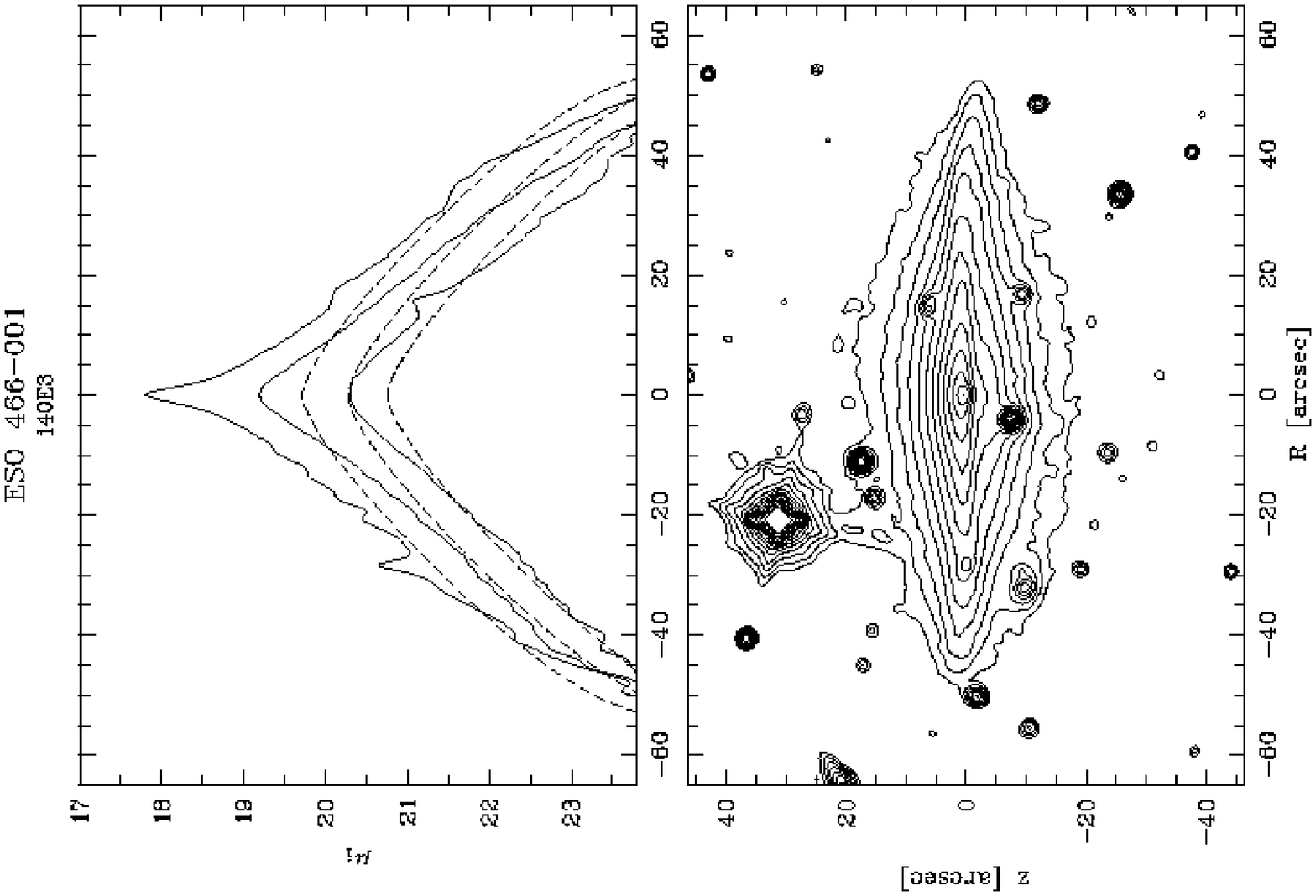,width=15cm,angle=90}
   \end{figure*}
\begin{figure*}
\vspace*{1cm}
\psfig{figure=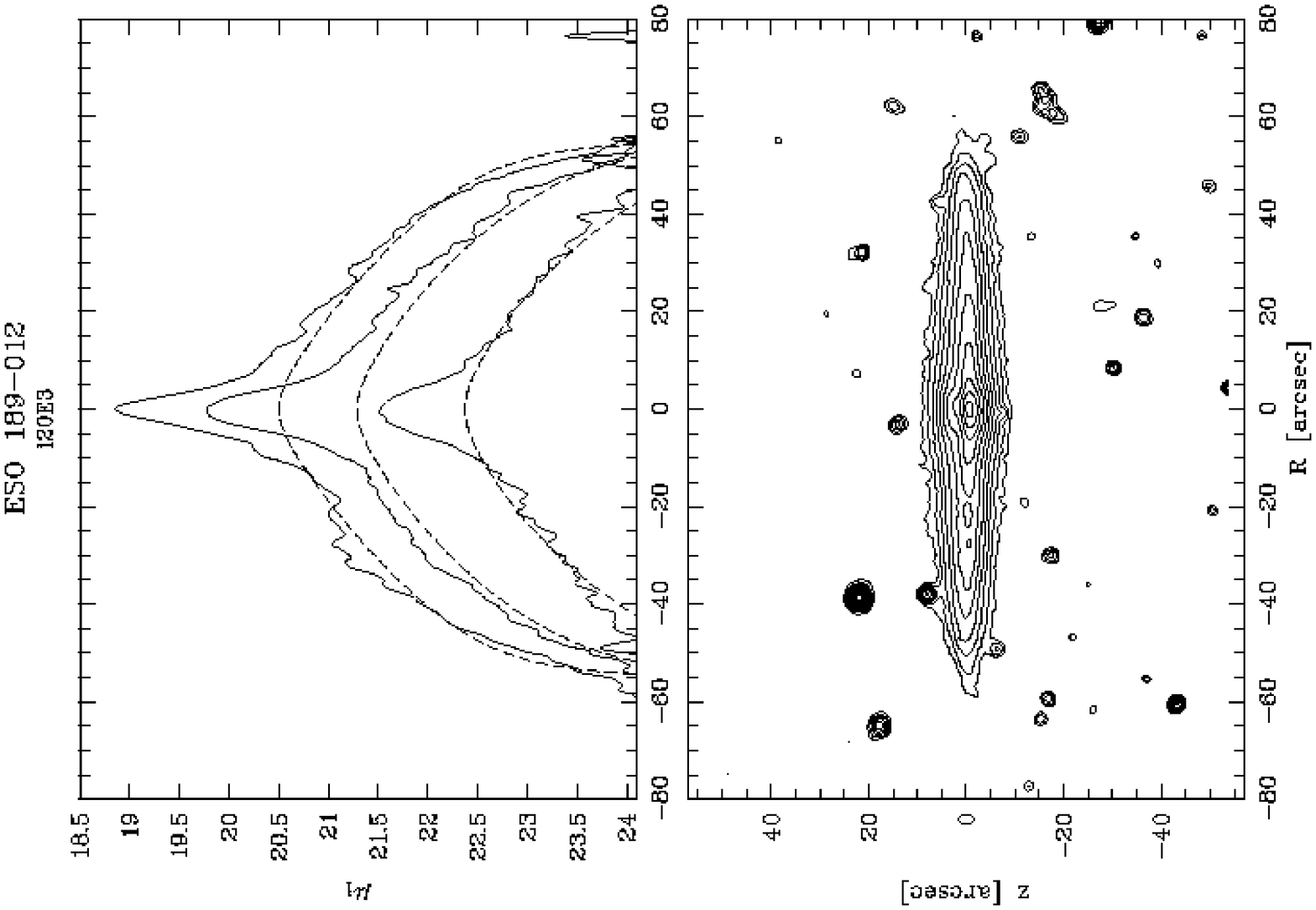,width=15cm,angle=90}
   \end{figure*}
\begin{figure*}
\psfig{figure=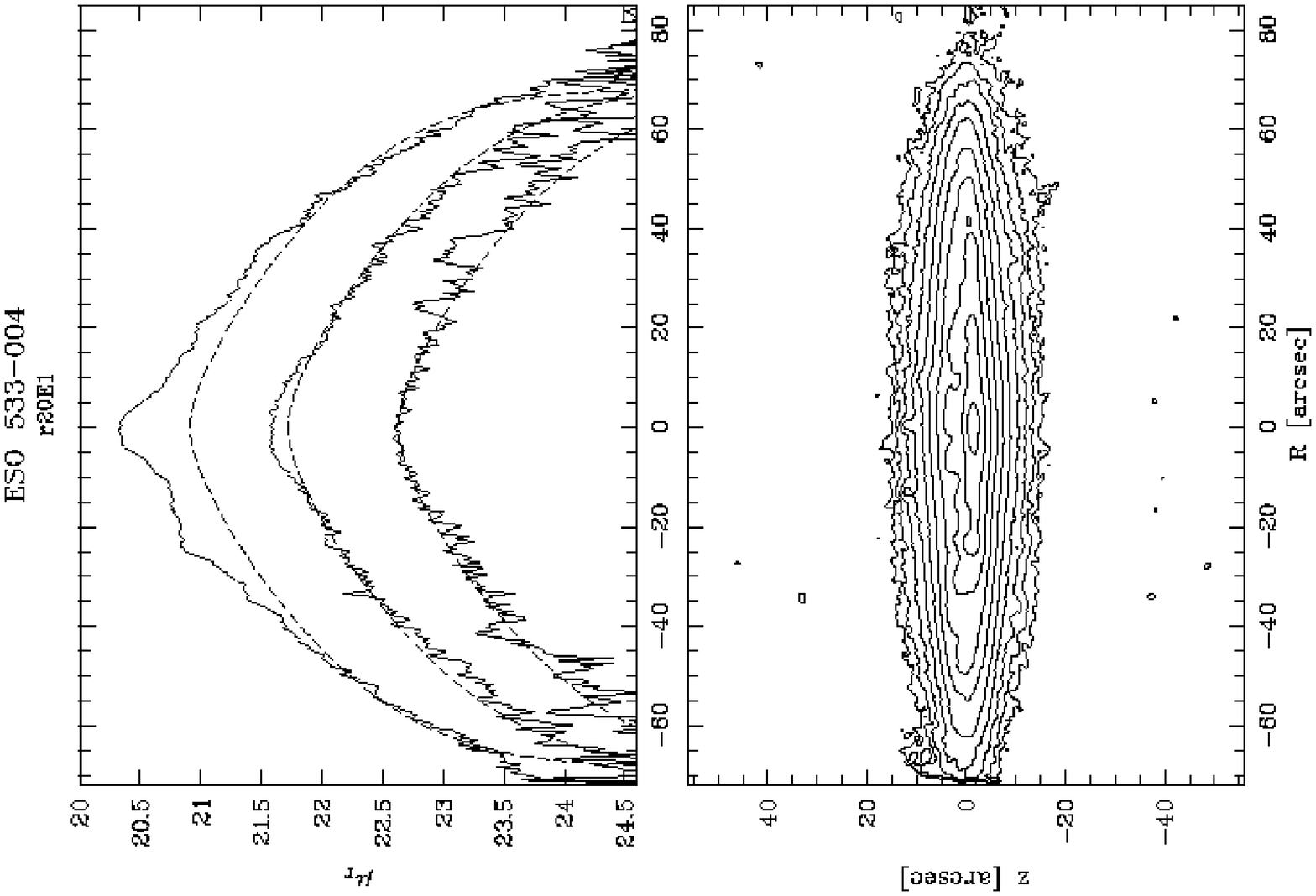,width=15cm,angle=90}
   \end{figure*}
\begin{figure*}
\vspace*{1cm}
\psfig{figure=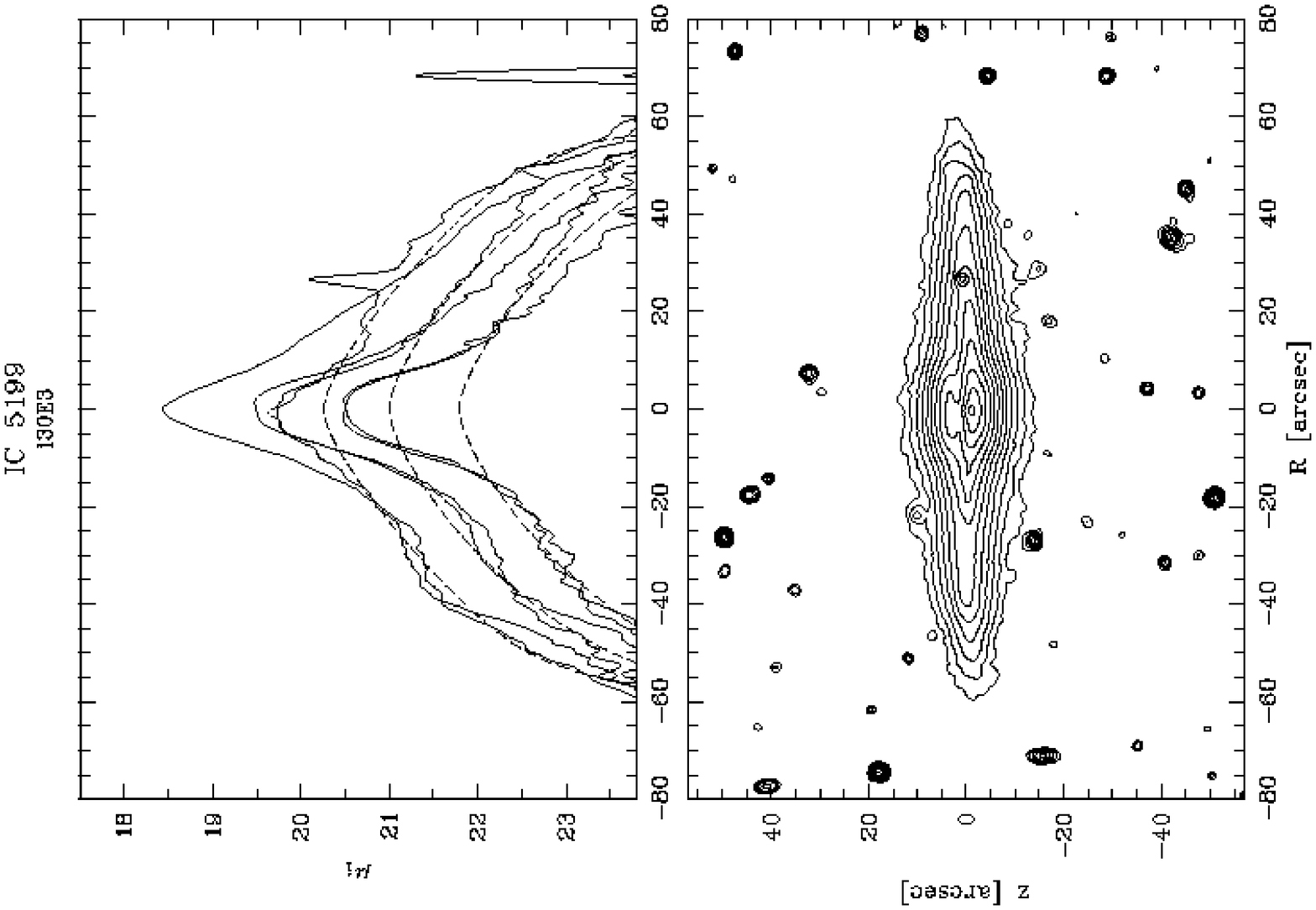,width=15cm,angle=90}
   \end{figure*}
\end{document}